\documentclass[aip,jcp,amsmath,amssymb,reprint]{revtex4-2}
\usepackage{mathptmx}
\usepackage[T1]{fontenc}
\usepackage[utf8]{inputenc}
\usepackage{mathtools}
\usepackage{amsmath}
\usepackage{graphicx}

\makeatletter

\providecommand{\tabularnewline}{\\}

\usepackage{dcolumn}
\usepackage{bm}
\usepackage{sidecap}
\usepackage{physics}

\usepackage{relsize}
\DeclareMathAlphabet{\mathcal}{OMS}{cmsy}{m}{n}
\DeclareFontFamily{OT1}{pzc}{}
\DeclareFontShape{OT1}{pzc}{m}{it}{<-> s * [1.100] pzcmi7t}{}
\DeclareMathAlphabet{\mathpzc}{OT1}{pzc}{m}{it}

\usepackage{multirow}
\usepackage{cleveref}
\allowdisplaybreaks

\makeatother

\begin{document}
\preprint{AIP/123-QED}
\title{Generalization of the Tavis-Cummings model for multi-level anharmonic
systems: insights on the second excitation manifold}
\author{J.Campos-Gonzalez-Angulo}
\affiliation{Department of Chemistry and Biochemistry, University of California
San Diego, La Jolla, California 92093, USA}

\author{J. Yuen-Zhou}
\email{joelyuen@ucsd.edu}

\homepage{http://yuenzhougroup.ucsd.edu}

\begin{abstract}
Confined electromagnetic modes strongly couple to collective excitations in ensembles of quantum emitters, producing light-matter hybrid states known as polaritons. Under such conditions, the discrete multilevel spectrum of molecular systems offers an appealing playground for exploring multiphoton processes. This work contrasts predictions from the Tavis-Cummings (TC) model, in which the material is a collection of two-level systems, with the implications of considering additional energy levels with harmonic and anharmonic structures. We discuss the exact eigenspectrum, up to the second excitation manifold, of an arbitrary number $N$ of oscillators collectively coupled to a single cavity mode in the rotating-wave approximation. Elaborating on our group-theoretic approach [\emph{New J. Phys.} 23, 063081 (2021)], we simplify the brute-force diagonalization of a gigantic $N^2\times N^2$ Hamiltonian (where $N=10^6-10^{10}$, as experiments suggest) to the diagonalization of, at most, $4\times4$ matrices. We thoroughly discuss the eigenstates and the consequences of weak and strong anharmonicities. Furthermore, we find resonant conditions between bipolaritons and anharmonic transitions where two-photon absorption can be enhanced. Finally, we conclude that energy shifts in the polaritonic states induced by anharmonicities become negligible for large $N$. Thus, calculations with a single or few emitters qualitatively fail to represent the nonlinear optical response of the collective strong coupling regime. Our work highlights the rich physics of multilevel anharmonic systems coupled to cavities absent in standard models of quantum optics. We also provide concise tabulated expressions for eigenfrequencies and transition amplitudes, which should serve as a reference for future spectroscopic studies of molecular polaritons.
\end{abstract}

\maketitle

\section{Introduction}

The strong coupling (SC) of confined photonic modes and excitations in
semiconductors and molecular materials leads to a wealth of interesting
chemical and condensed-matter physics phenomena \cite{RibeiroMartinez-MartinezDuEtAl2018,FeistGalegoGarcia-Vidal2018,FlickRiveraNarang2018,GarciaVidalFrancisco2021,Basov2021}
such as room-temperature Bose-Einstein condensation,\cite{Plumhof2014,Keeling2020,Zeb2020,PannirSivajothi2021}
long-range energy transfer,\cite{Georgiou2018,MartinezMartinez2019}
modification of chemical reactivity,\cite{Ebbesen2016,YuenZhou2019,HerreraOwrutsky2020}
and quantum information processing.\cite{Blais2004,Peng2021}. Strong
coupling leads to hybrid light-matter quantum modes known as polaritons,
originally characterized for electronic excitations in bulk materials,\cite{Agranovich1957,Hopfield1958} and later on studied in cavity systems.\cite{Weisbuch1992,Lidzey1998} Within linear response,
the formalism to understand these systems is naturally based on a
harmonic approximation for the material degrees of freedom.\cite{Hopfield1958}
However, to understand nonlinear optical properties, the material
degrees of freedom are typically modeled as two-level quantum emitters,
leading to the very well-studied Jaynes-Cummings and Rabi models in
the case of a single emitter,\cite{Rabi1937,JaynesCummings1963} or
the Tavis-Cummings (TC) and Dicke models in the collective regime.\cite{Dicke1954,TavisCummings1968}
Recently, there has been an increasing interest in the properties
of vibrational polariton systems resulting from the strong coupling
of infrared (IR) cavity modes and ensembles of localized high-frequency
vibrational modes in molecules in condensed phases.\cite{Crum2018,HiraiHutchisonUji-i2020,Erwin2021,Grafton2021,Nagarajan2021,Wang2021,Xiang2021,YuenZhou2021} The multilevel anharmonic spectrum
of these vibrational modes implies that their accurate description should
invoke more than two levels per emitter. Indeed, consideration of
more realistic vibrational SC systems involving
anharmonicities has been the subject of theoretical explorations,
such as the use of single-molecule models to explain cavity-induced
modifications to chemical reactivity,\cite{HernandezHerrera2019,TrianaHernandezHerrera2020,Fischer2021,Schaefer2021,Wang2021a} as well as the use
of many-molecule models to explain non-linear response experiments.\cite{Takemura2015,Xiang2018,Xiang2019,Autry2020,DelPo2020,Duan2021,Grafton2021}
In particular, recent work \cite{Ribeiro2018,SaurabhMukamel2018,Debnath2020,GuMukamel2020,Li2021,Ribeiro2021} has elucidated novel multiphoton absorption
phenomena where the TC description is insufficient, and a multilevel
anharmonic spectrum of the material is essential. These effects go beyond vibrational SC and should have analogues in other electromagnetic ranges, such as those systems under electronic SC where more than two electronic states per quantum emitter must be considered.

While theoretical complexities emerging from the multilevel anharmonic
spectrum of the oscillators are expected, they can be overcome by
taking advantage of the permutational symmetries derived from the
assumption that the emitters behave identically.\cite{Arecchi1972,Gilmore1972,GeggRichter2016,ShammahAhmedLambertEtAl2018}
For regimes in which the counterrotating (non-energy conserving) terms of the coupling
can be disregarded, the Hamiltonian can be separated according to
the total number of excitations allocated in the many-body states
that conform the basis of the Hilbert space.\cite{LeeLinksZhang2011,Skrypnyk2015}
The general strategy to simplify and solve the Schrödinger equation
for this instance can be found in ref. \onlinecite{CamposGonzalezAngulo2021}, work on which the present manuscript elaborates by describing
in detail the wavefunctions and the energy spectrum resulting from
diagonalizing the Hamiltonian of a collection of anharmonic multilevel
quantum emitters dipolarly coupled to a single cavity mode. In particular,
we focus on the structure of the subspace built from states bearing
two excitations and compare it with the well-known solutions of the
first excitation manifold. For systems under collective SC, the number
of emitters is very large at $N\approx10^{6}-10^{10}$, and a brute-force
numerical diagonalization to solve for the eigenspectrum of these
hybrid light-matter systems is unattainable. Our method exploiting
the permutational symmetry arising from the consideration of $N$
identical emitters dramatically reduces the problem to the diagonalization
of very small matrices.

The organization of this manuscript is as follows: in section \ref{sec:1man}
we revisit the first-excitation manifold and present the tools and vocabulary
of polaritonic states that permeate the remaining of the paper. Section
\ref{sec:2man} opens by introducing generalities of the states with
two excitations and proceeds with the presentation of the eigenstates
for the TC model. These solutions are contrasted with those obtained
after including an additional energy level per emitter, in the harmonic
and anharmonic regimes, respectively. Tabulated expressions are also
provided so that this work serves as a future reference for work on
molecular polariton spectroscopy. Finally, the conclusions are presented in section
\ref{sec:conc}.

\section{The first excitation manifold \label{sec:1man}}

For a collection of $N$ identical multi-level quantum emitters interacting
with a confined electromagnetic mode of frequency $\omega_0$ in the
regime where the rotating-wave approximation is valid, the total number
of excitations in the system, $n_{\textrm{exc}}$, is a conserved
quantity that defines the so-called excitation manifolds.

When $n_{\textrm{exc}}=0$, all the components of the ensemble
are in their respective ground states, and the system is characterized
by the ($N+1$)-body state $\ket{0}$.

A basis can be defined for states with $n_{\textrm{exc}}=1$, such
that the excitation is localized on each of the emitters. To be specific,
we have \begin{subequations}\label{eq:1xlocbas} 
\begin{align}
\ket{1_{0}}= & \hat{a}_{0}^{\dagger}\ket{0},\\
\intertext{and}\ket{1_{i>0}}= & \qty(\hat{\sigma}_{i}^{(0)})^\dagger\ket{0},
\end{align}
\end{subequations} where $\hat{a}_{0}^{\dagger}$ is the creation
operator acting on the EM mode, and $\hat{\sigma}_{i}^{(v)}$ is the
local operator acting on the $i$th emitter projecting the state $\ket{v+1}$
onto $\ket{v}$. The notation in \cref{eq:1xlocbas} implies that
all particles not explicitly indicated are in their respective ground
states.

The Hamiltonian $\hat{H}_{1}$ describes the system in the singly
excited manifold. In the basis defined in \cref{eq:1xlocbas}, the
matrix elements of $\hat{H}_{1}$, in units of $\hbar$, are given
by 
\begin{equation}
\mel{1_{i}}{\hat{H}_{1}}{1_{j}}=\begin{cases}
\omega_0 & \text{if }i=j=0\\
\omega_{10} & \text{if }i=j>0\\
g_{01} & \text{for }i=0,\enskip j>0\\
g_{10} & \text{for }i>0,\enskip j=0\\
0 & \text{otherwise}
\end{cases},\label{eq:h1}
\end{equation}
where $\omega_{uv}$ is the excitation frequency between molecular
energy levels $u$ and $v$, and $g_{uv}=\sqrt{\frac{\omega_{uv}}{2\hbar\epsilon_{0}\mathcal{V}}}\bra{u}\hat{\mu}\ket{v}$
is a coupling constant with $\mathcal{V}$ the mode volume, and $\hat{\mu}$
the transition dipole moment operator ($\hbar$ is the reduced Planck's constant, and $\epsilon_{0}$ is the permittivity of the vacuum).

The system remains unchanged upon permutations of the emitters; therefore,
its description can be simplified using the SU(2) collective operators\cite{BastarracheaMagnani2014,Choreno2018} \begin{subequations}
\begin{align}
\hat{J}_{-}^{(v)}= & \sum_{i=1}^{N}\hat{\sigma}_{i}^{(v)},\\
\hat{J}_{+}^{(v)}= & \qty(\hat{J}_{-}^{(v)})^{\dagger},\\
\intertext{and}\hat{J}_{0}^{(v)}= & \frac{1}{2}\qty[\hat{J}_{+}^{(v)},\hat{J}_{-}^{(v)}],
\end{align}
\end{subequations} which adhere to the angular momentum algebra,
i.e., \begin{subequations} 
\begin{align}
\qty[\hat{J}_{0}^{(v)},\hat{J}_{\pm}^{(v)}]= & \hat{J}_{\pm}^{(v)},\\
\qty[\hat{J}_{+}^{(v)},\hat{J}_{-}^{(v)}]= & 2\hat{J}_{0}^{(v)}.
\end{align}
\end{subequations} 
Since these operators are permutationally invariant, they carry the
totally-symmetric or trivial irreducible representation (irrep) of
the symmetric group $S_{N}$,\cite{Arecchi1972,Klimov2009} which will be denoted by A throughout
this manuscript.

The global ground state, $\ket{0}$, is also permutationally invariant;
therefore, the states $\ket{1_{0}}$ and 
\begin{equation}\label{eq:totsym1}
\begin{split}\ket{1_{\textrm{A}}}= & \frac{\hat{J}_{+}^{(0)}}{\sqrt{N}}\ket{0}\\
= & \frac{1}{\sqrt{N}}\sum_{i=1}^{N}\ket{1_{i}},
\end{split}
\end{equation}
carry the totally-symmetric irrep as well. On the other hand, the states 
\begin{equation}
\ket{1_{\textrm{B}(k)}}=\sum_{n=1}^{N}c_{n}^{(k)}\ket{1_{n}},\label{eq:drksts}
\end{equation}
where the coefficients $c_{n}^{(k)}$ fulfill $\sum_{n=1}^{N}c_{n}^{(k)}=0$
and $\sum_{n=1}^{N}c_{n}^{(k)*}c_{n}^{(k')}=\delta_{kk'}$, span the
$N-1$-dimensional Hilbert subspace corresponding to the standard
irrep of $S_{N}$, denoted by B, which is orthogonal to the totally-symmetric
one.

Since the subspace that the wavefunctions $\ket{1_{\textrm{B}(k)}}$
span is highly degenerate, the choice of basis is not unique.\cite{Strashko2016} The
most prominent examples of these bases are the Fourier basis: 
\begin{equation}
c_{n}^{(k)}=\frac{1}{\sqrt{N}}\exp\left(2\pi i\frac{(k-1)n}{N}\right),\label{eq:fourbas}
\end{equation}
in which $2\pi(k-1)/N$ is a wave vector, and the Schur-Weyl basis,
where 
\begin{equation}
c_{n}^{(k)}=\frac{\alpha_{n}^{(k)}}{\sqrt{k\qty(k-1)}},\label{eq:swbas}
\end{equation}
with 
\begin{equation}
\alpha_{n}^{(k)}=\begin{cases}
-1 & \text{for }1\leq n<k\\
k-1 & \text{if }n=k\\
0 & \text{for }k<n\leq N
\end{cases}.
\end{equation}

The states in the symmetry-adapted basis in \cref{eq:totsym1,eq:drksts} are also known as Dicke states,\cite{Dicke1954,Scully1997}
$\ket{J_{(0)},M_{J}^{(0)}}$, with quantum numbers given by the equations
\begin{subequations} 
\begin{align}
\hat{J}_{(0)}^{2}\ket{J_{(0)},M_{J}^{(0)}} & =J_{(0)}\qty(J_{(0)}+1)\ket{J_{(0)},M_{J}^{(0)}},\intertext{and}\hat{J}_{0}^{(0)}\ket{J_{(0)},M_{J}^{(0)}} & =M_{J}^{(0)}\ket{J_{(0)},M_{J}^{(0)}},
\end{align}
\end{subequations} where 
\begin{equation}
\hat{J}_{(0)}^{2}=\qty(\hat{J}_{0}^{(0)})^{2}+\frac{1}{2}\qty(\hat{J}_{+}^{(0)}\hat{J}_{-}^{(0)}+\hat{J}_{-}^{(0)}\hat{J}_{+}^{(0)}).
\end{equation}
For reasons that will become evident later, we found our labeling
of the irreps more convenient; however, the identification of our
notation with the corresponding quantum numbers from angular momentum
operators can be found in \cref{tab:symms}. 
\begin{table}
\caption{SU(2) and SU(3) identifiers of the molecular symmetry-adapted states.\label{tab:symms}}

\begin{ruledtabular}
\begin{tabular}{lllll}
State & $J_{0}$ & $M_{J}^{(0)}$ & $y$ & SU(3) multiplet\tabularnewline
\hline 
$\ket{0}$ & $N/2$ & $-N/2$ & $N/3$ & $\qty(N,0)$\tabularnewline
$\ket{1_{\textrm{A}}}$ & $N/2$ & $1-N/2$ & $N/3$ & $\qty(N,0)$\tabularnewline
$\ket{1_{\textrm{B}(k)}}$ & $N/2-1$ & $1-N/2$ & $N/3$ & $\qty(N-2,1)$\tabularnewline
$\ket{1_{\textrm{A}}^{2}}$ & $N/2$ & $2-N/2$ & $N/3$ & $\qty(N,0)$\tabularnewline
$\ket{1_{\textrm{B}(k)}^{2}}$ & $N/2-1$ & $2-N/2$ & $N/3$ & $\qty(N-2,1)$\tabularnewline
$\ket{1_{\textrm{C}(k\ell)}^{2}}$ & $N/2-2$ & $2-N/2$ & $N/3$ & $\qty(N-4,2)$\tabularnewline
$\ket{2_{\textrm{A}}}$ & $\qty(N-1)/2$ & $\qty(1-N)/2$ & $N/3-1$ & $\qty(N,0)$\tabularnewline
$\ket{2_{\textrm{B}(k)}}$ & $\qty(N-1)/2$ & $\qty(1-N)/2$ & $N/3-1$ & $\qty(N-2,1)$ \tabularnewline
\end{tabular}
\end{ruledtabular}

\end{table}

In the basis $\qty{\ket{1_0},\ket{1_\textrm{A}},\ket{1_{\textrm{B}(2)}},\ket{1_{\textrm{B}(3)}},\ldots,\ket{1_{\textrm{B}(N)}}}$, the Hamiltonian becomes
\begin{equation}\label{eq:h1sym}
\begin{split}
H_1=&\begin{pmatrix}
\omega_0&\sqrt{N}g_{01}&0&0&\cdots&0\\
\sqrt{N}g_{10}&\omega_{10}&0&0&\cdots&0\\
0&0&\omega_{10}&0&\cdots&0\\
0&0&0&\omega_{10}&\cdots&0\\
\vdots&\vdots&\vdots&\vdots&\ddots&\vdots\\
0&0&0&0&\cdots&\omega_{10}
\end{pmatrix}\\
=&\mathbf{H}_1^{(\textrm{A})}\oplus \qty(\mathbf{1}_{N-1}\otimes \mathbf{H}_1^{(\textrm{B})}),
\end{split}
\end{equation}
where $\mathbf{H}_1^{(\textrm{A})}$ is a $2\times2$ matrix, $\mathbf{H}_1^{(\textrm{B})}=\omega_{10}$, and $\mathbf{1}_{d}$ is the $d$-dimensional identity matrix.

From \cref{eq:h1sym}, it becomes clear that the dark states $\ket{1_{\textrm{B}(k)}}$
are eigenfunctions of $\hat{H}_{1}$. The other eigenstates are in the totally-symmetric sub-space:
\begin{equation}
\ket{1_{\pm}}=\pm h_{\pm}\ket{1_{0}}+h_{\mp}\ket{1_{\textrm{A}}},
\end{equation}
where $+$ and $-$ label the upper and lower polariton, respectively,
and $h_{\pm}=\sqrt{\frac{1}{2}\qty(1\pm\frac{\Delta}{\Omega_{10}})}$
are known as Hopfield coefficients. The corresponding eigenfrequencies
are 
\begin{equation}
\begin{split}\ev{\hat{H}_{1}}{1_{\pm}}= & \omega_{\pm}\\
= & \frac{\omega_0+\omega_{10}\pm\Omega_{10}}{2},
\end{split}
\end{equation}
where $\Omega_{10}=\sqrt{\Delta^{2}+4Ng_{01}^{2}}$ is the Rabi frequency,
and $\Delta=\omega_0-\omega_{10}$ is the detuning. Notice that the
use of permutational symmetry arguments to derive the analytical solution
of the eigenspectrum of $\hat{H}_{1}$ is a drastic simplification
of a brute-force numerical diagonalization of a $(N+1)$-dimensional matrix, which in the limit of $N=10^{6}-10^{10}$, as is the
case of collective SC, becomes intractable. We shall see how an analogous simplification
can be carried out for the second-excitation manifold.

The intensities of spectroscopic signals depend on the off-diagonal
matrix elements of the collective dipole operator $\hat{\boldsymbol{\mu}}=\sum_{i=1}^{N}{\hat{\mu}_{i}}$,
and the photon mode creation operator $\hat{a}_{0}^{\dagger}$. For
transitions between the ground state, $\ket{0}$, and the eigenstates
in the first excitation manifold we have 
\begin{align}
\bra{1_{\pm}}{\hat{\boldsymbol{\mu}}} & \ket{0}=\sqrt{N}h_{\mp}\mu_{10},\\
\bra{1_{\textrm{B}(k)}}{\hat{\boldsymbol{\mu}}} & \ket{0}=0,\label{eq:istand}\\
\bra{1_{\pm}}{\hat{a}_{0}^{\dagger}} & \ket{0}=\pm h_{\pm},\\
\intertext{and}\bra{1_{\textrm{B}(k)}}{\hat{a}_{0}^{\dagger}} & \ket{0}=0,
\end{align}
where $\mu_{10}=\mel{1}{\hat\mu}{0}$ is the dipole moment
for the $0\to1$ transition of the bare emitters. \Cref{eq:istand} can be understood under the consideration that
the operator $\hat{\boldsymbol{\mu}}$ is totally-symmetric, and the
states $\ket{1_{\textrm{B}(k)}}$ carry an orthogonal irrep.

\section{The second excitation manifold \label{sec:2man}}

The states with two quanta can be of the form 
\begin{equation}
\ket{1_{i}1_{j}}=\qty(\hat{\sigma}_{j\neq i}^{(0)})^\dagger\ket{1_{i}},
\end{equation}
in which the $i$th and the $j$th particle are both in their first
excited state, or \begin{subequations}\label{eq:2xlocbas} 
\begin{align}
\ket{2_{0}}= & \frac{\hat{a}_{0}^{\dagger}}{\sqrt{2}}\ket{1_{0}},\\
\intertext{and}\ket{2_{i>0}}= & \qty(\hat{\sigma}_{i}^{(1)})^\dagger\ket{1_{i}},
\end{align}
\end{subequations} where the $i$th particle is in its second excited
state. As in the singly excited manifold, all particles not explicitly
indicated are in their respective ground state. Notice that a brute-force
numerical diagonalization of the second-excitation manifold Hamiltonian
$\hat{H}_{2}$ would require diagonalization of ${N+2 \choose 2}$-dimensional matrices, that is unattainable for the large number $N$ of
emitters which concern us in the context of collective SC. Instead,
as in the previous section, we will exploit the permutational symmetry
arising from the consideration of identical emitters coupled to the
cavity.

Just as before, a symmetry-adapted basis can be defined such that
there are wavefunctions carrying the totally-symmetric irrep: $\ket{2_{0}}$,
\begin{equation}
\begin{split}\ket{1_{0}1_{\textrm{A}}}= & \hat{a}_{0}^{\dagger}\ket{1_{\textrm{A}}}\\
= & \frac{\hat{J}_{+}^{(0)}}{\sqrt{N}}\ket{1_{0}},
\end{split}
\end{equation}
\begin{equation}
\begin{split}\ket{1_{\textrm{A}}^{2}}= & \frac{\hat{J}_{+}^{(0)}\ket{1_{\textrm{A}}}}{\sqrt{2\qty(N-1)}}\\
= & \sqrt{\frac{2}{N(N-1)}}\sum_{i=1}^{N-1}\sum_{j=i+1}^{N}\ket{1_{i}1_{j}},
\end{split}
\end{equation}
and 
\begin{equation}
\begin{split}\ket{2_{\textrm{A}}}= & \hat{J}_{+}^{(1)}\ket{1_{\textrm{A}}}\\
= & \frac{1}{\sqrt{N}}\sum_{i=1}^{N}\ket{2_{i}}.
\end{split}
\label{eq:2a}
\end{equation}
There are also wavefunctions carrying the standard irrep: 
\begin{equation}
\ket{1_{0}1_{\textrm{B}(k)}}=\hat{a}_{0}^{\dagger}\ket{1_{\textrm{B}(k)}},
\end{equation}
\begin{equation}
\begin{split}\ket{1_{\textrm{B}(k)}^{2}}= & \frac{\hat{J}_{+}^{(0)}}{\sqrt{N-2}}\ket{1_{\textrm{B}(k)}}\\
= & \sum_{m=1}^{N-1}\sum_{n=m+1}^{N}c_{mn}^{(k)}\ket{1_{m}1_{n}},
\end{split}
\end{equation}
and 
\begin{equation}
\begin{split}\ket{2_{\textrm{B}(k)}}= & \hat{J}_{+}^{(1)}\ket{1_{\textrm{B}(k)}}\\
= & \sum_{n=1}^{N}c_{n}^{(k)}\ket{2_{n}},
\end{split}
\label{eq:2b}
\end{equation}
where the coefficients $c_{n}^{(k)}$ are the same as in \cref{eq:fourbas,eq:swbas},
and the coefficients $c_{mn}^{(k)}=\qty(c_{m}^{(k)}+c_{n}^{(k)})/\sqrt{N-2}$
fulfill \begin{subequations} 
\begin{align}
\sum_{m=1}^{N-1}\sum_{n=m+1}^{N}c_{mn}^{(k)}= & 0,\\
\intertext{and}\sum_{m=1}^{N-1}\sum_{n=m+1}^{N}c_{mn}^{(k)}c_{mn}^{(k')*}= & \delta_{kk'}.
\end{align}
\end{subequations} In the Fourier basis, these coefficients are 
\begin{equation}
c_{mn}^{(k)}=\frac{2\exp\left[\pi i(k-1)\frac{m+n}{N}\right]}{\sqrt{N(N-2)}}\cos\left(\pi(k-1)\frac{m-n}{N}\right),
\end{equation}
while in the Schur-Weyl basis, 
\begin{equation}
c_{mn}^{(k)}=\frac{\alpha_{mn}^{(k)}}{\sqrt{(N-2)k(k-1)}},
\end{equation}
where 
\begin{equation}
\alpha_{mn}^{(k)}=\begin{cases}
-2 & \text{for }1\leq m<k,\enskip1<n<k\\
k-2 & \text{for }1\leq m<k,\enskip n=k\\
-1 & \text{for }1\leq m<k,\enskip k<n\leq N\\
k-1 & \text{if }m=k,\enskip k<n\leq N\\
0 & \text{otherwise}
\end{cases}.
\end{equation}
Finally, there are wavefunctions, 
\begin{equation}
\ket{1_{\textrm{C}(k\ell)}^{2}}=\sum_{m=1}^{N-1}\sum_{n=m+1}^{N}c_{mn}^{(k\ell)}\ket{1_{m}1_{n}},
\end{equation}
carrying the $N(N-3)/2$-dimensional irrep that is orthogonal to both
the totally-symmetric and the standard irreps. Their coefficients fulfill \begin{subequations}
\begin{align}
\sum_{m=1}^{N-1}\sum_{n=m+1}^{N}c_{mn}^{(k\ell)}= & 0,\\
\sum_{m=1}^{N-1}\sum_{n=m+1}^{N}c_{mn}^{(k\ell)}c_{mn}^{(k')*}= & 0,\intertext{and}\sum_{m=1}^{N-1}\sum_{n=m+1}^{N}c_{mn}^{(k\ell)}c_{mn}^{(k'\ell')*}= & \delta_{kk'}\delta_{\ell\ell'}.
\end{align}
\end{subequations} These coefficients, in the Fourier basis, have
the general form 
\begin{equation}
c_{mn}^{(k\ell)}=\sum_{q}\sum_{K(q)}\psi(K_{k\ell},q_{k\ell})\textrm{e}^{\pi iK_{k\ell}\frac{m+n}{N}}\cos\left(\pi q_{kl}\frac{m-n}{N}\right),\label{eq:4drk}
\end{equation}
where $K_{k\ell}$ labels a center-of-mass wave number, $q_{k\ell}$
identifies a relative wave number, and $\psi(q_{k\ell})$ is a coefficient
resulting from symmetrization. The structure of this basis is discussed
in greater detail in appendix \ref{sec:four2x}. In the Schur-Weyl
basis, the coefficients are given by 
\begin{equation}
c_{mn}^{(k\ell)}=-\frac{\alpha_{mn}^{(k\ell)}}{\sqrt{k(k-1)(\ell-2)(\ell-3)}},
\end{equation}
with 
\begin{equation}
\alpha_{mn}^{(k\ell)}=\begin{cases}
-2 & \text{for }1\leq m<k,1<n<k\\
k-2 & \text{for }1\leq m<k,n=k\\
-1 & \text{for }1\leq m<k,k<n<\ell\\
\ell-3 & \text{for }1\leq m<k,n=\ell\\
k-1 & \text{for }m=k,k<n<\ell\\
-(k-1)(\ell-3) & \text{if }m=k,n=\ell\\
0 & \text{otherwise}
\end{cases}.
\end{equation}

We note that the involvement of $\hat{J}_{+}^{(1)}$ in \cref{eq:2a,eq:2b}
implies that the SU(2) algebra with operators $\hat{J}_{x}^{(0)}$
is insufficient to characterize the states with two excitations in
the same emitter.\cite{Cordero2013} These states need labels from the SU(3) algebra
for their correct identification. In \cref{tab:symms}, we present
these labels, including the eigenvalues, $y$, of the operator 
\begin{equation}
\hat{Y}=\frac{2}{3}\qty(\qty[\hat{J}_{+}^{(1)},\hat{J}_{-}^{(1)}]-\hat{J}_{0}^{(0)}),
\end{equation}
usually known as hypercharge in the literature of particle physics.\cite{Bohr1998}
We also require labels for the SU(3) irrep or multiplet; the methods
to identify them can be found elsewhere.\cite{Lipkin2002} As we can see, $J_{0}$,
is useful to identify the symmetry of states with at most one excitation
per emitter, but is not capable of classifying the remaining states.
On the other hand, the irreps of SU(3) thoroughly fulfill this job.
However, keeping track of that many quantum numbers is cumbersome
for the purposes of this work, and the A, B, C scheme here introduced
conveys the important information in a more condensed way.

Up to this point, we know that the matrix element 
\begin{equation}
\mel{\Psi_{\Gamma}}{\hat{H}_{2}}{\Psi_{\Gamma'}}=0,
\end{equation}
if $\Gamma\neq\Gamma'$, where $\hat{H}_{2}$ is the Hamiltonian with
$n_{\textrm{exc}}=2$, and $\ket{\Psi_{\Gamma}}$ is a symmetrized
state carrying the irrep $\Gamma$ which can be either A, B or C.

The form of the Hamiltonian in the subspaces carrying the irreps A
and B depends on the considered spectrum of the emitters,
and will be discussed in the following sections. On the other hand,
the states $\ket{1_{\textrm{C}(k\ell)}^{2}}$ do not couple to the
EM mode in any capacity nor among themselves, and are therefore eigenstates
of $\hat{H}_{2}$ in every model discussed below, i.e., 
\begin{equation}
\mel{1_{\textrm{C}(k\ell)}^{2}}{\hat{H}_{2}}{1_{\textrm{C}(k'\ell')}^{2}}=2\omega_{10}\delta_{kk'}\delta_{\ell\ell'};
\end{equation}
as such, they will not be discussed in the next subsections.

\begin{figure}
\includegraphics[width=0.4\textwidth]{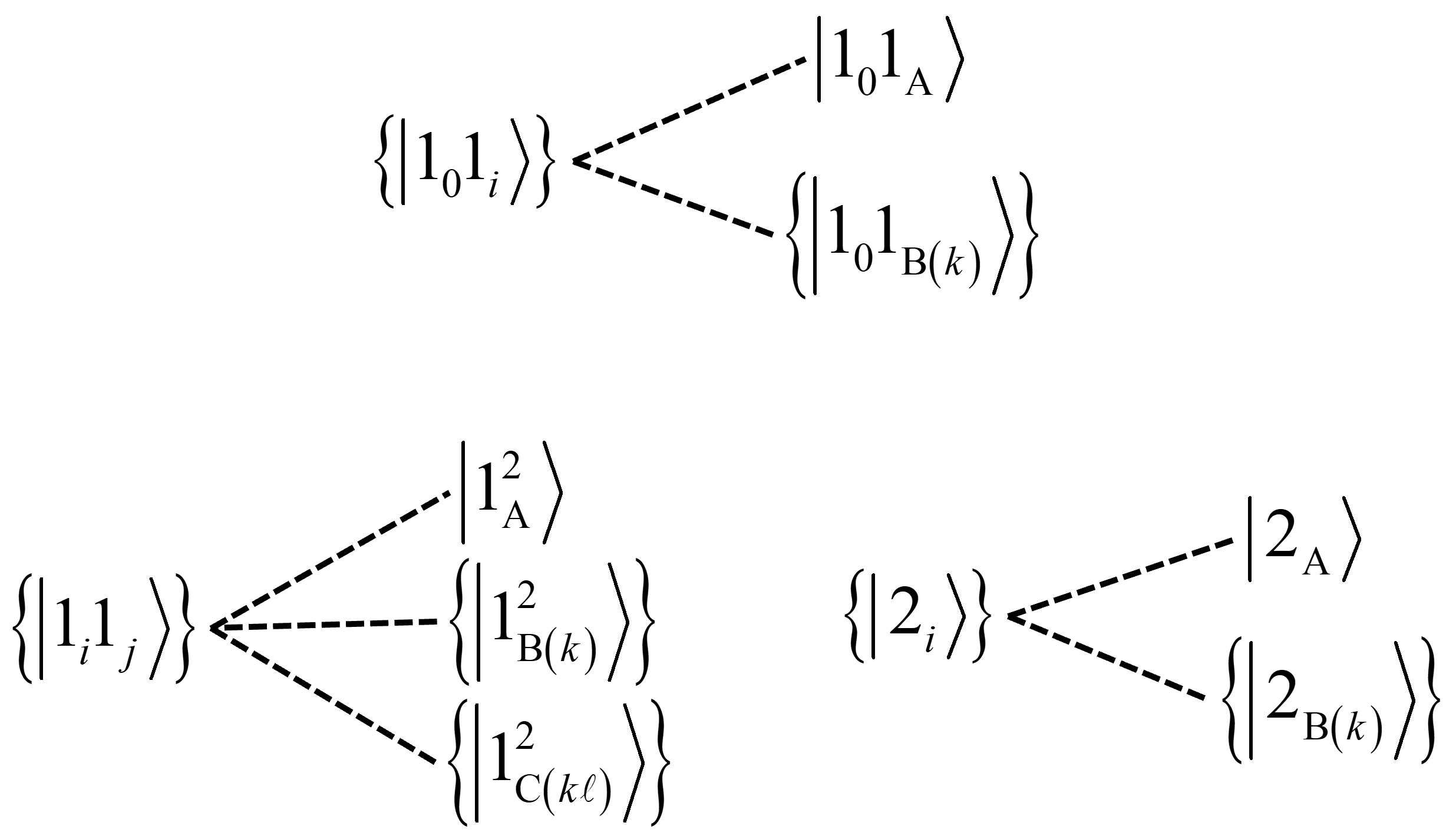} \caption{Distribution of states in the doubly excited manifold among irreps.\label{fig:splits}}
\end{figure}

\Cref{fig:splits} illustrates the distribution of every spectral configuration among the irreps. To be specific, the Hamiltonian matrix in the second excitation manifold decomposes according to
\begin{equation}
\mathbf{H}_2=\mathbf{H}_2^{(\textrm{A})}\oplus\qty(\mathbf{1}_{N-1}\otimes \mathbf{H}_2^{(\textrm{B})})\oplus \qty(\mathbf{1}_{N\qty(N-3)/2}\otimes\mathbf{H}_2^{(\textrm{C})}),
\end{equation}
where the dimensions of $\mathbf{H}_2^{(\textrm{A})}$ and $\mathbf{H}_2^{(\textrm{B})}$ depend upon the particular energetic structure of the emitters, and $\mathbf{H}_2^{(\textrm{C})}=2\omega_{10}$.

\subsection{Tavis-Cummings}

If the emitters are well-approximated by two-level systems, the matrix
elements of the Hamiltonian for the doubly excited manifold are
\begin{subequations}
\begin{align}
\bra{2_{0}}{\hat{H}_{2}} & \ket{2_{0}}=2\omega_0,\\
\bra{2_{0}}{\hat{H}_{2}} & \ket{1_{i}1_{j}}=\sqrt{2}g_{01}\delta_{i0},\\
\bra{1_{i}1_{j}}{\hat{H}_{2}} & \ket{2_{0}}=\sqrt{2}g_{10}\delta_{i0},\\
\intertext{and}\bra{1_{i}1_{j}}{\hat{H}_{2}} & \ket{1_{i'}1_{j'}}=
\begin{cases}
\omega_0+\omega_{10} & \text{if }i=i'=0,\enskip j=j'\\
2\omega_{10} & \text{if }i=i'\neq0,\enskip j=j'\neq0\\
g_{01} & \text{if }i=0\neq i',\enskip j=j'\\
g_{10} & \text{if }i\neq0=i',\enskip j=j'\\
0 & \text{otherwise}
\end{cases}.
\end{align}
\end{subequations}
In turn, the matrix in the totally-symmetric subspace, i.e., with the symmetrized basis $\qty{\ket{2_0},\ket{1_0 1_\textrm{A}},\ket{1_\textrm{A}^2}}$, is
\begin{equation}\label{eq:tca} 
\mathbf{H}_{2,\textrm{TC}}^{(\textrm{A})}=
\begin{pmatrix}
2\omega_0&\sqrt{2N}g_{10}&0\\
\sqrt{2N}g_{01}&\omega_0+\omega_{10}&\sqrt{2\qty(N-1)}g_{10}\\
0&\sqrt{2\qty(N-1)}g_{01}&2\omega_{10}.
\end{pmatrix}
\end{equation}
For states carrying the standard irrep, $\qty{\ket{1_0 1_{\textrm{B}(k)}},\ket{1_{\textrm{B}(k)}^2}}$, the corresponding Hamiltonian matrix is
\begin{equation}
\mathbf{H}_{2,\textrm{TC}}^{(\textrm{B})}=
\begin{pmatrix}
\omega_0+\omega_{10}&\sqrt{N-2}g_{10}\\
\sqrt{N-2}g_{01}&2\omega_{10}.
\end{pmatrix}
\end{equation}
Notice that, as opposed to the singly excited manifold, the sub-space carrying
the standard irrep of the doubly excited Hamiltonian is no longer
diagonal in the symmetry-adapted basis.

Diagonalization of $\mathbf{H}_{2,\textrm{TC}}^{(\textrm{A})}$ gives rise to the so-called bipolaritonic states\cite{Ivanov1995,Takemura2015,DelPo2020,Ribeiro2021}
\begin{subequations}
\begin{align}
{\hat{H}_{2}} & \ket{2_{\pm}}=\omega_{2\pm}\ket{2_{\pm}},\\
\intertext{and}{\hat{H}_{2}} & \ket{1_{+}1_{-}}=\omega_{1_{+}1_{-}}\ket{1_{+}1_{-}}.
\end{align}
\end{subequations} Eigenstates and eigenfrequencies of this $3\times3$
matrix can be easily obtained numerically. Analytical expressions,
however, are cumbersome (see appendix \ref{sec:TCexact}) unless the
reasonable approximation $N\gg1$ is considered, in which case: 
\begin{align}
\ket{2_{\pm}}_{\textrm{TC}}= & h_{\pm}^{2}\ket{2_{0}}\pm\sqrt{2}h_{+}h_{-}\ket{1_{0}1_{\textrm{A}}}+h_{\mp}^{2}\ket{1_{\textrm{A}}^{2}},\label{eq:2tc}\\
\intertext{and}\ket{1_{+}1_{-}}_{\textrm{TC}}= & \sqrt{2}h_{+}h_{-}\qty(-\ket{2_{0}}+\ket{1_{\textrm{A}}^{2}})\nonumber \\
 & +\qty(h_{+}^{2}-h_{-}^{2})\ket{1_{0}1_{\textrm{A}}}.
\end{align}

The eigenvectors of $\mathbf{H}_{2,\textrm{TC}}^{(\textrm{B})}$ give the coefficients for the oftentimes referred to as ``upper(lower) polariton-dark states'': \cite{Takemura2015,DelPo2020}
\begin{equation}
\ket{1_{\pm}1_{\textrm{B}(k)}}_{\textrm{TC}}=\pm h_{\pm}\ket{1_{0}1_{\textrm{B}(k)}}+h_{\mp}\ket{1_{\textrm{B}(k)}^{2}},\label{eq:11btc}
\end{equation}
which fulfill
\begin{equation}
\hat{H}_2\ket{1_{\pm}1_{\textrm{B}(k)}}_{\textrm{TC}}=\omega_{1_{\pm}1_{\textrm{B}}}\ket{1_{\pm}1_{\textrm{B}(k)}}_{\textrm{TC}}.
\end{equation} 

The eigenfrequencies are tabulated in \cref{tab:freqs}, while the
matrix elements of transition operators between the eigenstates in
the first excitation manifold and those in \crefrange{eq:2tc}{eq:11btc}
are included in \cref{tab:intensitiesA,tab:intensitiesB,tab:intensitiesA0,tab:intensitiesB0}.
The analytical expressions for the solutions with arbitrary $N$ can
be found in Appendix \ref{sec:TCexact}.

\subsection{Harmonic limit}

When two levels are not enough to depict the structure of the emitters,
the simplest way to add complexity to the system is by considering
the emitters as harmonic oscillators, i.e., with evenly-spaced levels
in their energy spectrum. This assumption is valid for modes related
to bonds with high dissociation energies. In this scenario, the new
light-matter Hamiltonian incorporates additional matrix elements given
by \begin{subequations} 
\begin{align}
\bra{2_{i>0}}{\hat{H}_{2}} & \ket{2_{j}}=2\omega_{10}\delta_{ij},\\
\bra{1_{0}1_{i}}{\hat{H}_{2}} & \ket{2_{j}}=\sqrt{2}g_{01}\delta_{ij},\\
\intertext{and}\bra{2_{i}}{\hat{H}_{2}} & \ket{1_{0}1_{j}}=\sqrt{2}g_{10}\delta_{ij}.
\end{align}
\end{subequations} Notice that the harmonic approximation assumes
$\omega_{12}=\omega_{10}$ and $g_{12}=\sqrt{2}g_{01}$.

With the new symmetrized contributions to the space of functions, namely $\ket{2_\textrm{A}}$ for $\mathbf{H}_2^{(\textrm{A})}$, and $\qty{\ket{2_{\textrm{B}(k)}}}$ for $\mathbf{H}_2^{(\textrm{B})}$, the expanded Hamiltonian matrices are
\begin{multline}\label{eq:hoa}
\mathbf{H}_{2,\textrm{HO}}^{(\textrm{A})}=\\
\begin{pmatrix}
2\omega_0&\sqrt{2N}g_{10}&0&0\\
\sqrt{2N}g_{01}&\omega_0+\omega_{10}&\sqrt{2\qty(N-1)}g_{10}&\sqrt{2}g_{10}\\
0&\sqrt{2\qty(N-1)}g_{01}&2\omega_{10}&0\\
0&\sqrt{2}g_{01}&0&2\omega_{10}
\end{pmatrix},
\end{multline}
for the totally-symmetric irrep, and
\begin{equation}\label{eq:hob}
\mathbf{H}_{2,\textrm{HO}}^{(\textrm{B})}=
\begin{pmatrix}
\omega_0+\omega_{10}&\sqrt{N-2}g_{10}&g_{21}\\
\sqrt{N-2}g_{01}&2\omega_{10}&0\\
g_{12}&0&2\omega_{10}
\end{pmatrix},
\end{equation}
for the standard irrep.

While the separation of the Hamiltonian accomplished with this approach
significantly facilitates the numerical diagonalization, a simpler
approach to obtain analytical results can be achieved by noticing
that a harmonic bi-linear Hamiltonian can always be written as a sum
of normal harmonic modes. In other words, the Hamiltonian of the system
becomes 
\begin{equation}
\begin{split}
\hat{H}_{\textrm{HO}}=&\omega_0\hat{a}_{0}^{\dagger}\hat{a}_{0}+\omega_{10}\sum_{i=1}^{N}\hat{a}_{i}^{\dagger}\hat{a}_{i}\\
&+\sum_{i=1}^{N}\qty(g_{10}\hat{a}_{i}^{\dagger}\hat{a}_{0}+g_{01}\hat{a}_{0}^{\dagger}\hat{a}_{i}),
\end{split}
\end{equation}
where $\hat{a}_{i}=\sum_{v=0}^{\infty}\sqrt{v+1}\hat{\sigma}_{i}^{(v)}$
is the bosonic annihilation operator acting on the $i$th emitter. A set
of symmetrized operators can be defined such that the Hamiltonian
becomes 
\begin{equation}
\begin{split}\hat{H}_{\textrm{HO}}= & \omega_0\hat{a}_{0}^{\dagger}\hat{a}_{0}+\omega_{10}\hat{a}_{\textrm{A}}^{\dagger}\hat{a}_{\textrm{A}}+\sqrt{N}\qty(g_{10}\hat{a}_{\textrm{A}}^{\dagger}\hat{a}_{0}+g_{01}\hat{a}_{0}^{\dagger}\hat{a}_{\textrm{A}})\\
 & +\omega_{10}\sum_{k=2}^{N}\hat{a}_{\textrm{B}(k)}^{\dagger}\hat{a}_{\textrm{B}(k)},
\end{split}
\end{equation}
with creation operators $\hat{a}_{\textrm{A}}^{\dagger}=\sum_{v=0}^{\infty}\sqrt{v+1}\hat{J}_{+}^{(v)}/\sqrt{N}$,
and $a_{\textrm{B}(k)}^{\dagger}=\sum_{n=1}^{N}c_{n}^{(k)}a_{i}^{\dagger}$,
where the coefficients $c_{n}^{(k)}$ are the same as in \cref{eq:drksts}.
Furthermore, defining the polaritonic modes through the creation operators
$\hat{a}_{\pm}^{\dagger}=\pm h_{\pm}\hat{a}_{0}^{\dagger}+h_{\mp}\hat{a}_{\textrm{A}}^{\dagger}$,
allows to rewrite the Hamiltonian as 
\begin{equation}
\hat{H}_{\textrm{HO}}=\omega_{+}\hat{a}_{+}^{\dagger}\hat{a}_{+}+\omega_{-}\hat{a}_{-}^{\dagger}\hat{a}_{-}+\omega_{10}\sum_{k=2}^{N}\hat{a}_{\textrm{B}(k)}^{\dagger}\hat{a}_{\textrm{B}(k)}.
\end{equation}
Since the modes defined by this Hamiltonian are independent, any combination
of creation operators applied to the groundstate will generate an
eigenstate. Consequently, for the totally-symmetric subspace, \begin{subequations}
\begin{align}
\ket{2_{\pm}}_{\textrm{HO}}= & \frac{\qty(\hat{a}_{\pm}^{\dagger})^{2}}{\sqrt{2}}\ket{0}\nonumber \\
= & h_{\pm}^{2}\ket{2_{0}}\pm\sqrt{2}h_{+}h_{-}\ket{1_{0}1_{\textrm{A}}}\\
 & +h_{\mp}^{2}\qty(\frac{\sqrt{N-1}\ket{1_{\textrm{A}}^{2}}+\ket{2_{\textrm{A}}}}{\sqrt{N}}),\nonumber \\
\ket{1_{+}1_{-}}_{\textrm{HO}}= & \hat{a}_{\pm}^{\dagger}\hat{a}_{\mp}^{\dagger}\ket{0}\nonumber \\
= & -\sqrt{2}h_{+}h_{-}\ket{2_{0}}+\qty(h_{+}^{2}-h_{-}^{2})\ket{1_{0}1_{\textrm{A}}}\\
 & +\sqrt{2}h_{+}h_{-}\qty(\frac{\sqrt{N-1}\ket{1_{\textrm{A}}^{2}}+\ket{2_{\textrm{A}}}}{\sqrt{N}})\nonumber \\
\intertext{and}\ket{2_{D\textrm{A}}}_{\textrm{HO}}= & \frac{1}{\sqrt{N-1}}\sum_{k=2}^{N}\frac{\qty(\hat{a}_{\textrm{B}(k)}^{\dagger})^{2}}{\sqrt{2}}\ket{0}\nonumber \\
= & \frac{-\ket{1_{\textrm{A}}^{2}}+\sqrt{N-1}\ket{2_{\textrm{A}}}}{\sqrt{N}}.
\end{align}
\end{subequations}

The eigenstates carrying the standard representation are \begin{subequations}
\begin{align}
\ket{1_{\pm}1_{\textrm{B}(k)}}_{\textrm{HO}}= & \hat{a}_{\pm}^{\dagger}\hat{a}_{\textrm{B}(k)}\ket{0}=\hat{a}_{\textrm{B}(k)}^{\dagger}\hat{a}_{\pm}\ket{0}\nonumber \\
= & \pm h_{\pm}\ket{1_{0}1_{\textrm{B}(k)}}\\
 & +h_{\mp}\qty(\frac{\sqrt{N-2}\ket{1_{\textrm{B}(k)}^{2}}+\sqrt{2}\ket{2_{\textrm{B}(k)}}}{\sqrt{N}})\nonumber \\
\intertext{and}\ket{2_{D\textrm{B}(k)}}_{\textrm{HO}}= & \frac{-\sqrt{2}\ket{1_{\textrm{B}(k)}^{2}}+\sqrt{N-2}\ket{2_{\textrm{B}(k)}}}{\sqrt{N}}.
\end{align}
\end{subequations} The states $\ket{2_{D\textrm{A}}}$ and $\ket{2_{D\textrm{B}(k)}}$ are purely molecular and
account for the addition of states of the form $\ket{2_{i}}$ to the
Hilbert space.

\begin{figure*}
\includegraphics[width=1\textwidth]{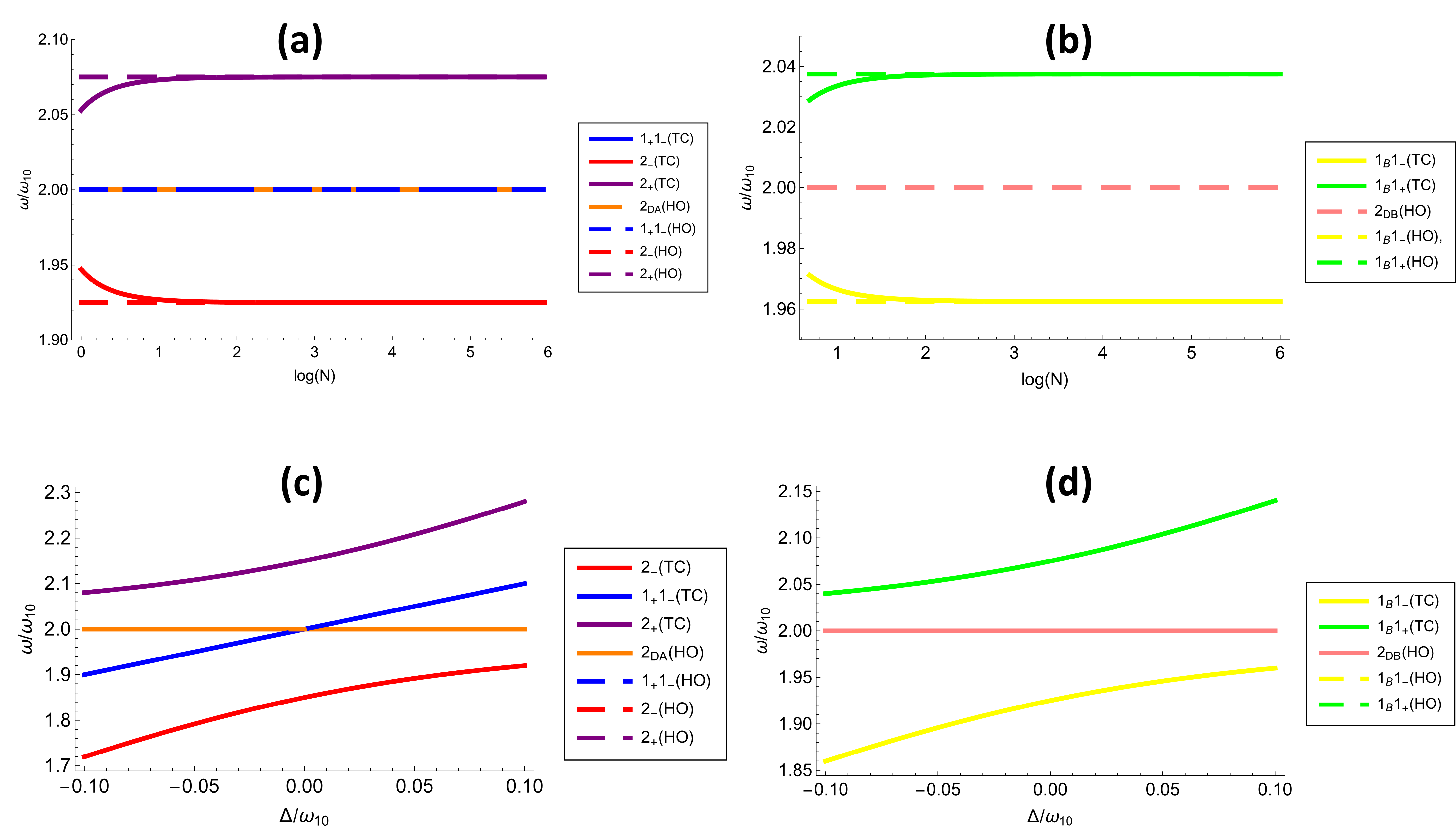} \caption{Energy spectrum of the doubly excited manifold in the Tavis-Cummings (TC)
and the harmonic oscillator (HO) models. Spectra of totally-symmetric states
(a) and states carrying the standard irrep (b) as a function of the
number of emitters, $N$, for a cavity in resonance with the $0\to1$
vibrational transition. Spectra of totally-symmetric states (c) and
states carrying the standard irrep (d) as a function of the detuning
between the cavity and the $0\to1$ transition, for $N=10^{9}$. All
plots were calculated with constant collective coupling amplitude
of $7\omega_{10}/100$. Notice that (b) starts at $\log(N)>0$ because
there are no states carrying the standard representation for $N=1$.
Also, notice that the $2_{D\mathrm{A}}$ and $2_{D\mathrm{B}}$ states are only defined for
the HO model, as the TC model does not afford states
with two excitations in the same emitter. Finally, the lack of photonic
character for $2_{D\mathrm{A}}$ and $2_{D\mathrm{B}}$ implies that
they remain dispersionless as a function of detuning $\Delta$.\label{fig:tcho}}
\end{figure*}

The emerging eigenvalues are on display in \cref{tab:freqs}, and
the transition intensities between first and second excitation manifolds
for this model are summarized in \cref{tab:intensitiesA,tab:intensitiesB,tab:intensitiesA0,tab:intensitiesB0}.
\Cref{fig:tcho} compares the behavior of the energy spectrum for
the TC model and the harmonic approximation as functions of the number
of coupled emitters and detuning. As it can be seen, the TC model
and the harmonic approximation yield indistinguishable results as
the number of emitters increases. To explain this observation, we remark that the interaction between the cavity and the $1\to2$ transition in the emitters is
not extensive, as opposed to the $0\to1$ transition. Therefore, although the overall light-matter coupling in the multi-level system is stronger than in the TC model,
the contribution from the additional level gets diluted as the ensemble grows. We emphasize again
the emergence of states in the multi-level case, $\ket{2_{D\textrm{A}}}$ and $\ket{2_{D\textrm{B}}}$,
whose lack of photonic character makes them impervious to detuning.

\subsection{General (anharmonic) case.}

More realistically, emitters deviate from the behavior of harmonic
oscillators. In particular, for molecular vibrations, there is a mechanical anharmonicity reflecting
that vibrational energy levels are not evenly spaced, i.e., 
\begin{equation}
\omega_{12}=\omega_{10}\left(1-\chi\right),
\end{equation}
where $\chi$ is the mechanical anharmonicity constant. Additionally, an electrical
anharmonicity constant, $\gamma$, can also be defined stemming from 
\begin{equation}
g_{12}=\sqrt{2}g_{01}\left(1+\gamma\right).
\end{equation}
These anharmonicities are generally present in electronic transtions as well. Notice also that the TC model effectively corresponds to the anharmonic case with $\chi=0$ and $\gamma=-1$.

In terms of these parameters, the matrix elements of the light-matter
Hamiltonian need to be updated to \begin{subequations} 
\begin{align}
\bra{2_{i>0}}{\hat{H}_{2}} & \ket{2_{j}}=\qty(\omega_{10}+\omega_{21})\delta_{ij},\\
\bra{1_{0}1_{i}}{\hat{H}_{2}} & \ket{2_{j}}=g_{12}\delta_{ij},\\
\bra{2_{i}}{\hat{H}_{2}} & \ket{1_{0}1_{j}}=g_{21}\delta_{ij}.
\end{align}
\end{subequations}

\begin{figure*}
\includegraphics[width=0.9\textwidth]{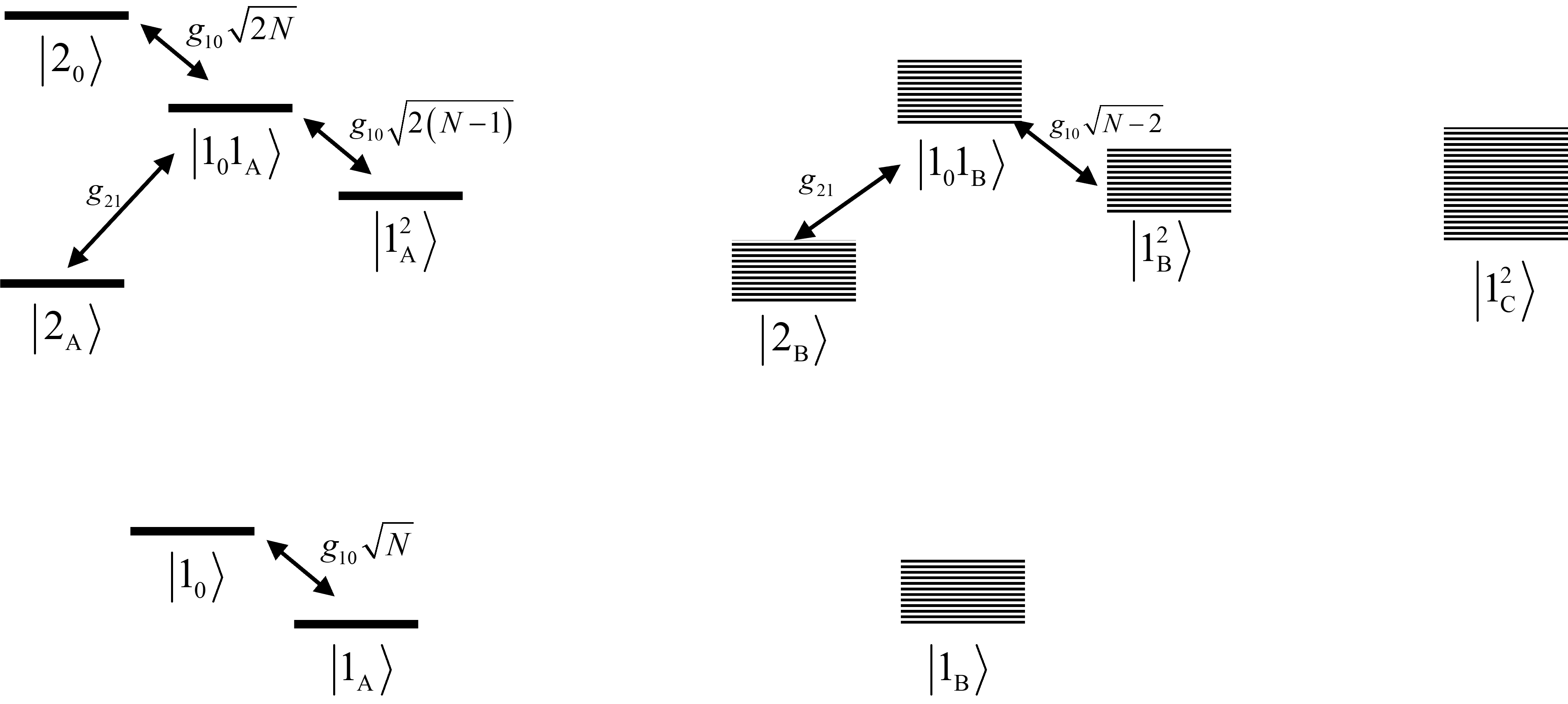} \caption{Interaction scheme among levels in the singly and doubly excited manifolds.
States carrying the totally-symmetric (A) irreducible representation
are non-degenerate, while states with the standard (B) and C irreps
are $(N-1)$-fold and $N(N-3)/2$-fold degenerate, respectively. States
with labels $2_{\textrm{A}}$ and $2_{\textrm{B}}$ do not appear
in the Tavis-Cummings model.\label{fig:ladlevs}}
\end{figure*}

In the symmetry-adapted basis they become \begin{subequations}
\begin{align}
\bra{2_{\textrm{A}}}{\hat{H}_{2}} & \ket{2_{\textrm{A}}}=\omega_{10}+\omega_{21},\\
\bra{2_{\textrm{A}}}{\hat{H}_{2}} & \ket{1_{0}1_{\textrm{A}}}=g_{12},\\
\intertext{and}\bra{1_{0}1_{\textrm{A}}}{\hat{H}_{2}} & \ket{2_{\textrm{A}}}=g_{21},
\end{align}
\end{subequations} for the totally-symmetric irrep, and \begin{subequations}
\begin{align}
\bra{2_{\textrm{B}(k)}}{\hat{H}_{2}} & \ket{2_{\textrm{B}(k')}}=\qty(\omega_{10}+\omega_{21})\delta_{kk'},\\
\bra{2_{\textrm{B}(k)}}{\hat{H}_{2}} & \ket{1_{0}1_{\textrm{B}(k')}}=g_{12}\delta_{kk'},\\
\intertext{and}\bra{1_{0}1_{\textrm{B}(k)}}{\hat{H}_{2}} & \ket{2_{\textrm{B}(k')}}=g_{21}\delta_{kk'},
\end{align}
\end{subequations} for the standard irrep. The effects of anharmonicities can thus be introduced as corrections to the harmonic Hamiltonian matrices in \cref{eq:hoa,eq:hob}. Explicitly, the full Hamiltonian matrices, carrying the irrep $\Gamma$, in the second-excitation manifold read
\begin{equation}
\mathbf{H}_2^{(\Gamma)}=\mathbf{H}_{2,\textrm{HO}}^{(\Gamma)}+{\mathbf{H}'_{2}}^{(\Gamma)},
\end{equation}
where
\begin{equation}
{\mathbf{H}'_2}^{(\textrm{A})}=
\begin{pmatrix}
0&0&0&0\\
0&0&0&\gamma\sqrt{2}g_{10}\\
0&0&0&0\\
0&\gamma\sqrt{2}g_{01}&0&-\chi\omega_{10}
\end{pmatrix}
\end{equation}
and
\begin{equation}
{\mathbf{H}'_2}^{(\textrm{B})}=
\begin{pmatrix}
0&0&\gamma\sqrt{2}g_{10}\\
0&0&0\\
\gamma\sqrt{2}g_{01}&0&-\chi\omega_{10}
\end{pmatrix},
\end{equation}
are the anharmonic corrections with $\Gamma=\textrm{A}$ and B, respectively. The interaction between
levels according to their symmetry is summarized in \cref{fig:ladlevs}.

\begin{figure*}
\includegraphics[width=1\textwidth]{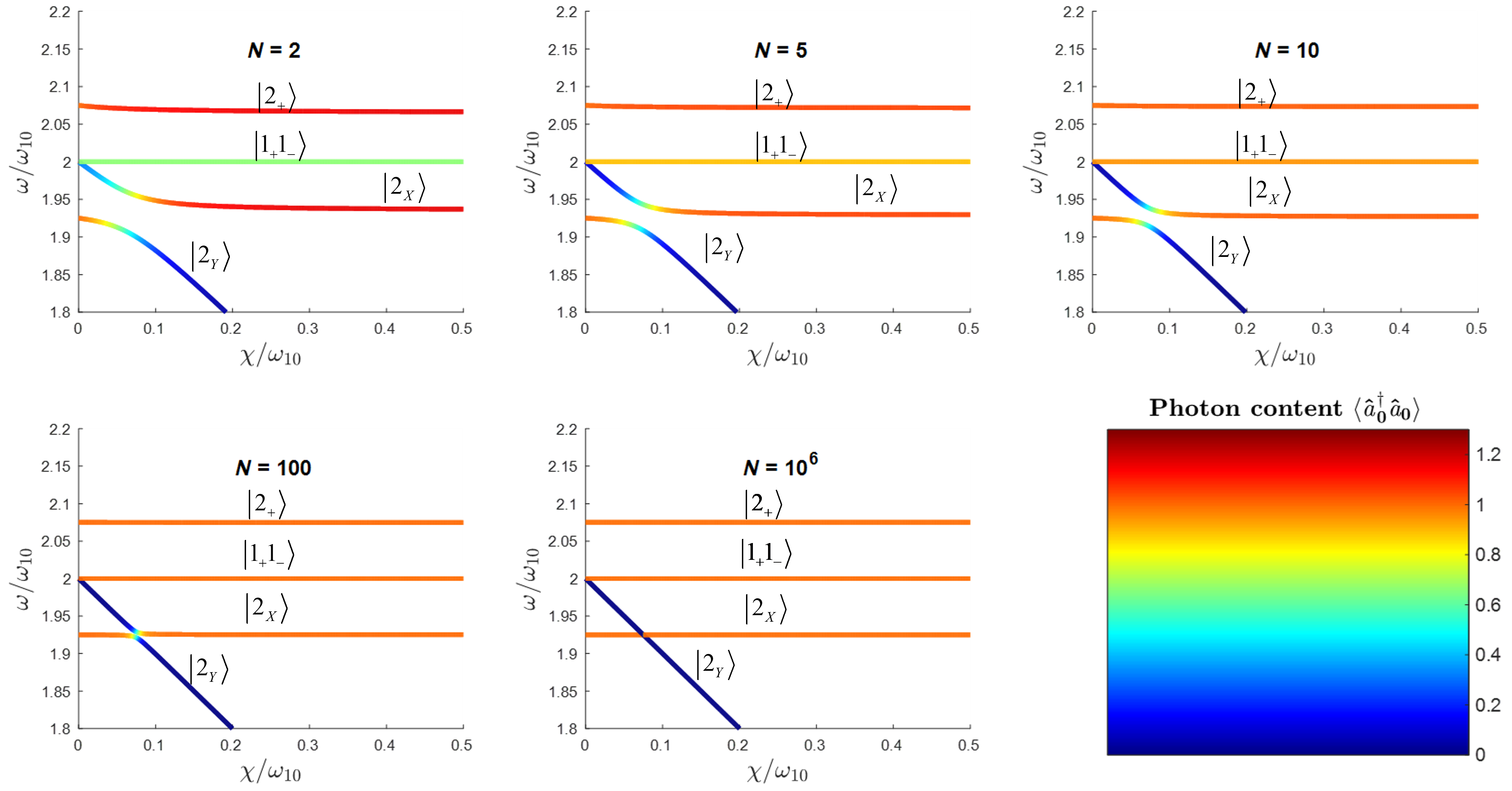} \caption{Exact energy spectrum of doubly-excited totally-symmetric eigenstates
as a function of anharmonicity, $\chi$, for several numbers of Morse
emitters. Resonant case ($\omega_0=\omega_{10}$). The labeling identifies the frequencies of the adiabatic eigenstates as a function of $\chi/\omega_{10}$. Since the states $\ket{2_{+}}$ and $\ket{1_{+}1_{-}}$ are well-separated in energy, they have been labeled to reflect that they are very similar to the corresponding harmonic eigenstates. The $\ket{2_X}$ and $\ket{2_Y}$ adiabatic eigenstates show avoided crossings resulting from anharmonic mixing of the harmonic states $\ket{2_{D\textrm{A}}}$ and $\ket{2_{-}}$. In all cases, $\ket{2_{X}}\approx\ket{2_{D\textrm{A}}}$
and $\ket{2_{Y}}\approx\ket{2_{-}}$ for $\chi\to0$, and $\ket{2_{X}}\approx\ket{2_{-}}$
and $\ket{2_{Y}}\approx\ket{2_{D\textrm{A}}}$ for large $\chi$. The photon content
is evaluated as the expectation value of the number operator $\hat{a}_{0}^{\dagger}\hat{a}_{0}$.
Notice that, for large $N$, the avoided crossings giving rise to the
$\ket{2_{X}}$ and $\ket{2_{Y}}$ states are essentially crossings
between the very anharmonic  state $\ket{2_{D\mathrm{A}}}$ and the bipolariton $\ket{2_{-}\}}$. These crossings are loci of two-photon absorption enhancement.\label{fig:specad0}}
\end{figure*}

To analyze the effect of anharmonicity in the energy spectra we consider, without loss of generality,
the emitters as Morse oscillators, i.e., the mass-normalized normal-mode
elongation, $x$, is subject to a potential energy of the form\cite{Morse1929} 
\begin{equation}
V(x)=\hbar\omega_{10}\frac{\qty(1+\chi)^{2}}{2\chi}\qty[1-\exp(-\sqrt{\frac{\omega_{10}\chi}{\hbar}}x)]^{2}.
\end{equation}
For this system, the electrical and mechanical anharmonicities are
related through\cite{Lima2005}
\begin{equation}
\gamma=\frac{1}{2}\sqrt{\frac{(1+\chi)(16-\chi^{2})}{2(2+\chi)}}-1.
\end{equation}
\Cref{fig:specad0} compares the exact energy spectra of totally-symmetric states
as a function of anharmonicity for several values of $N$ obtained
through numerical diagonalization of $\mathbf{H}_2^{(\textrm{A})}$. 

\begin{SCfigure*}
\caption{Analogous plot to \cref{fig:specad0}, panel $N=10^{6}$, except that the cavity is negatively detuned ($\omega_{0}=0.95\omega_{10}$). The eigenstate $\ket{2_{+}}$ is well-separated in energy and has been labeled to reflect that it is very similar to the corresponding harmonic eigenstate. The $\ket{2_R}$, $\ket{2_S}$, $\ket{2_T}$ adiabatic eigenstates show weakly (because of large $N$) avoided crossings resulting from anharmonic mixing of harmonic states $\ket{2_{D\textrm{A}}}$, $\ket{1_{+}1_{-}}$, and $\ket{2_{-}}$. Three anharmonicity
regimes can be recognized: small $\chi$ ($\ket{2_{R}}\approx\ket{2_{D\textrm{A}}}$,
$\ket{2_{S}}\approx\ket{1_{+}1_{-}}$, and $\ket{2_{T}}\approx\ket{2_{-}}$),
intermediate $\chi$ ($\ket{2_{R}}\approx\ket{1_{+}1_{-}}$, $\ket{2_{S}}\approx\ket{2_{D\textrm{A}}}$,
and $\ket{2_{T}}\approx\ket{2_{-}}$), and large $\chi$ ($\ket{2_{R}}\approx\ket{1_{+}1_{-}}$,
$\ket{2_{S}}\approx\ket{2_{-}}$, and $\ket{2_{T}}\approx\ket{2_{D\textrm{A}}}$). The avoided crossings occur at resonances between the very anharmonic $\ket{2_{D\textrm{A}}}$ with the bipolaritons $\ket{1_{+}1_{-}}$ and $\ket{2_{-}}$, the latter of which features a significant photonic content. These crossings are the loci of two-photon absorption enhancement. Notice that, owing to the negative detuning, there are two (rather than just one) such crossings, as opposed to the resonant case shown in \cref{fig:specad0}.\label{fig:negdet}}
\includegraphics[width=0.5\textwidth]{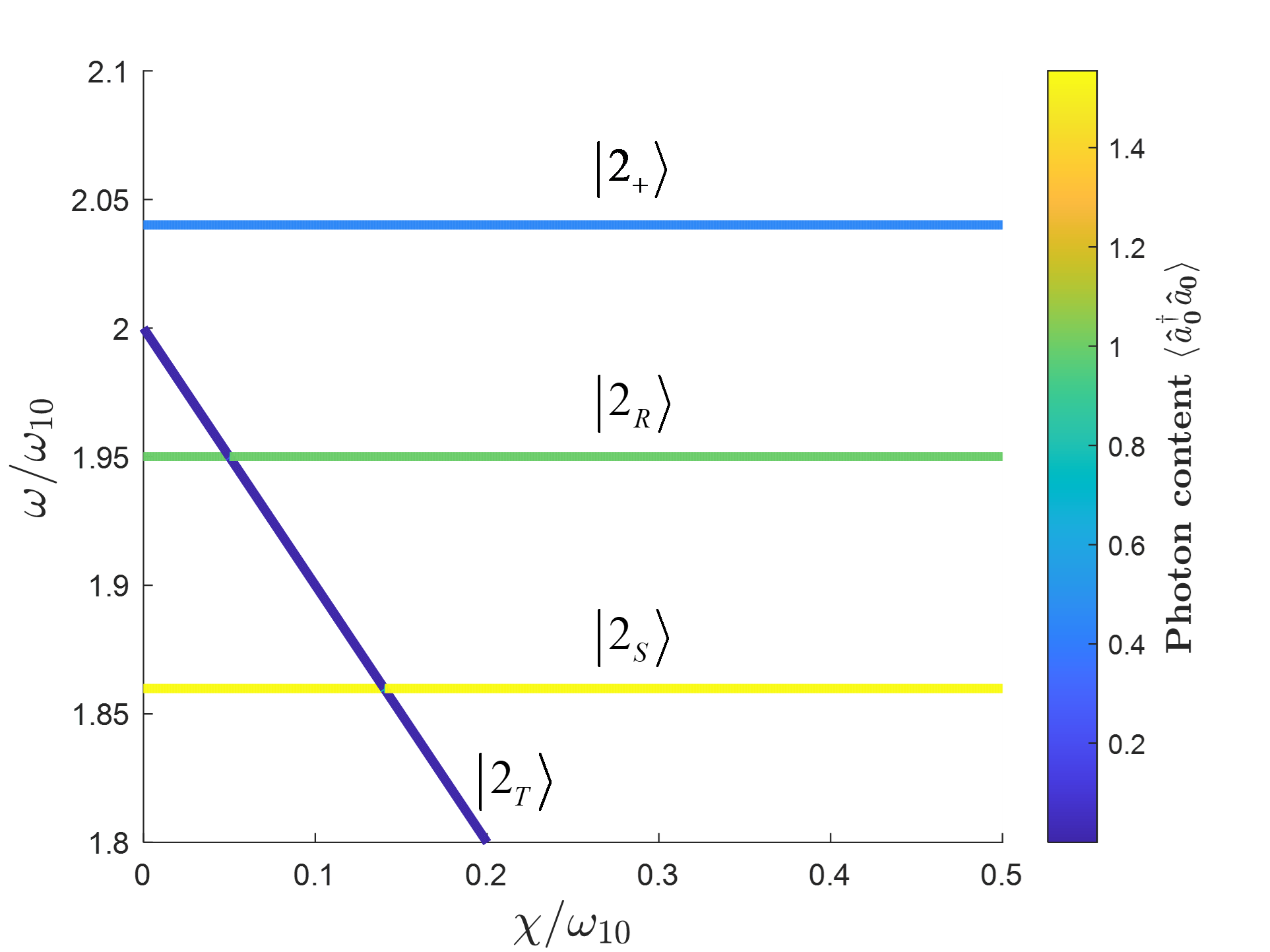}
\end{SCfigure*}

\begin{figure*}
\includegraphics[width=\textwidth]{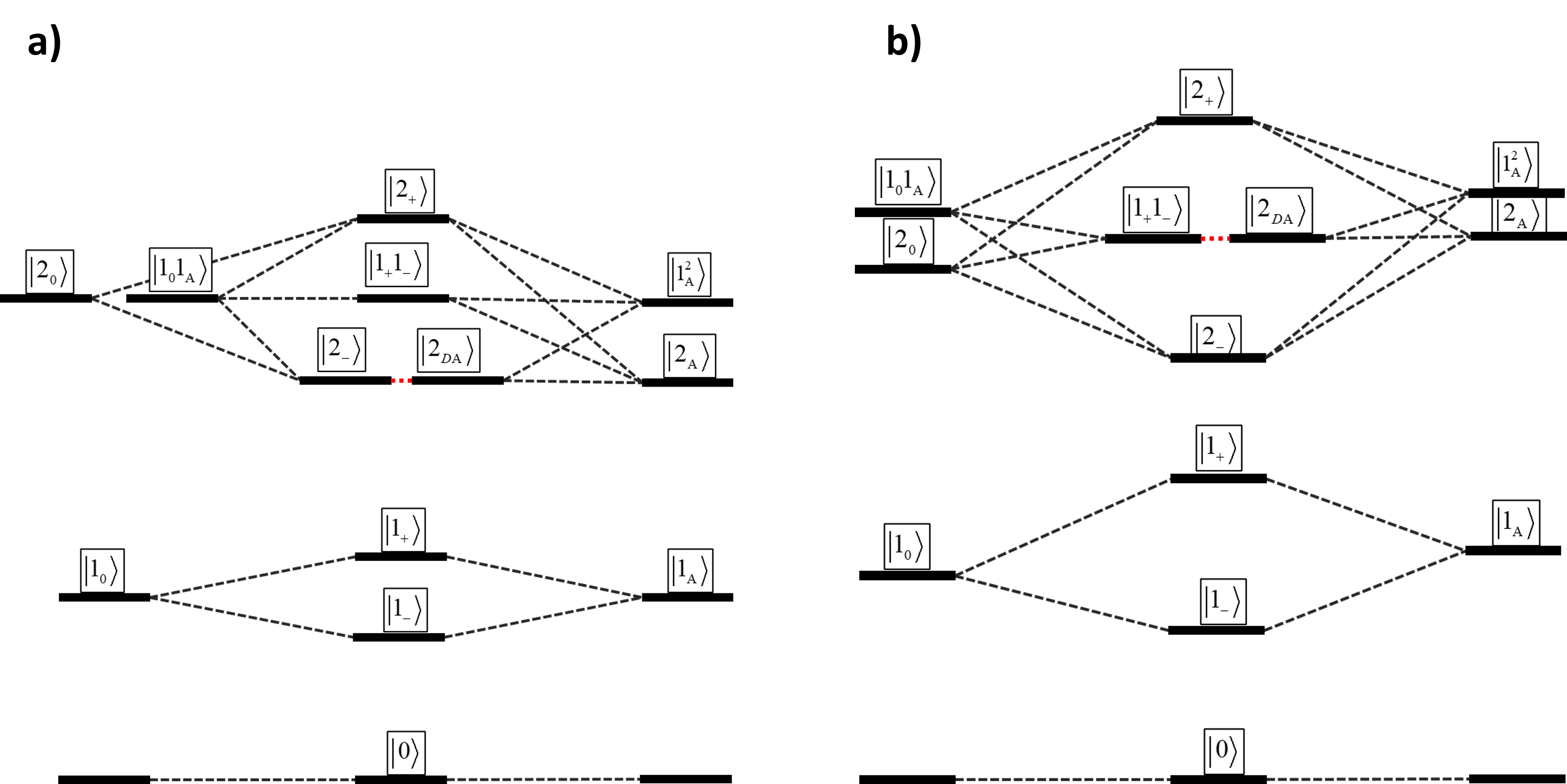} \caption{Correlation diagrams of polaritonic states under the conditions that afford enhanced TPA. (a) If the photon mode is resonant with the $0\to1$ molecular transition (i.e. $\omega_{0}=\omega_{10}$), enhancement of TPA occurs when $\omega_{2-}=\omega_{10}+\omega_{21}$ due to anharmonic coupling between the bipolariton $\ket{2_{-}}$ and the anharmonic state $\ket{2_{D\textrm{A}}}$ (red dotted line). (b) If the photon mode is negatively detuned from the $0\to1$ molecular transition (i.e. $\omega_{0}<\omega_{10}$), enhancement of TPA can also occur when $\omega_{1+1-}=\omega_{10}+\omega_{21}$ due to anharmonic coupling between the bipolariton $\ket{1_{+}1_{-}}$ and the anharmonic state $\ket{2_{D\textrm{A}}}$ (red dotted line).  \label{fig:corrdiag}}
\end{figure*}

As can be seen, the harmonic states $\ket{2_{-}}$ and $\ket{2_{D\textrm{A}}}$
interact anharmonically with a coupling strength determined by the single-molecule
light-matter coupling constant $g_{21}$. Therefore, it is possible
to find a set of parameters in the Hamiltonian for which these states
are near-resonant for large $N$, which is the limit that concerns
us for collective SC. Under this condition, the absorption of two
photons with the frequency of the $\ket{0}\to\ket{1_{-}}$ transition
will experience an enhancement, as discussed in refs. \onlinecite{Xiang2019,Ribeiro2021}.
This phenomenon can be understood as follows: the bipolariton $\ket{2_{-}}$
provides an optical window to funnel energy efficiently into the anharmonic
$\ket{2_{D\textrm{A}}}$ state.\cite{Li2021,Ribeiro2021} The effect can be observed when $\omega_{2-}\approx2\omega_{10}-\chi$. At light-matter resonance with the fundamental transition ($\omega_{0}=\omega_{10}$), this condition translates into the more intuitive $\Omega_{10}\approx\chi$, i.e.,  the
Rabi splitting must be tuned to to match the anharmonic shift. Note
that the bipolariton $\ket{2_{-}}$ is the only possible optical window
to resonate with $\ket{2_{D\textrm{A}}}$ when $\omega_{0}=\omega_{10}$.
However, if the photon is negatively detuned from the fundamental
transition, $\omega_{0}<\omega_{10}$, a second bipolariton $\ket{1_{+}1_{-}}$
can also be in resonance with $\ket{2_{D\textrm{A}}}$, as shown in
\cref{fig:negdet}. This effect was not discussed in Ref. \onlinecite{Ribeiro2021},
and should provide an extra tuning parameter to enhance multiphoton
absorption processes in polaritonic systems. The conditions for enhanced two-photon absorption (TPA) are better illustrated in the correlation diagrams of \cref{fig:corrdiag}.

The calculations above show that, as opposed to the case of linear response, simulations
with small $N$ are not a reliable representation of the nonlinear optics of systems under collective SC
 for anharmonicities up to one order of magnitude larger than the Rabi splitting, even if the intensity of the total coupling, 
$\sqrt{N}g_{10}$, is fixed to match experimental values of $\Omega_{10}$. Thus, we believe that the interpretation of experimental spectra as reported in refs. \onlinecite{Grafton2021,Duan2021} using a single-molecule model is worth revisiting.

While the numerical results above shed much light on the physics of
the problem for all ranges of anharmonicity, it is iluminating to carry
out a perturbation theory analysis for small values of $\chi$ to obtain closed approximate
expressions for eigenvalues and eigenvectors of $\hat{H}_{2}$. To
zeroth order, we have $\ket{\Psi}^{(0)}=\ket{\Psi}_{\textrm{HO}}$.
The first-order corrections to the eigenfrequencies of the HO limit
are included in \cref{tab:freqs}. The $O(\chi,\gamma)$ corrections
to the doubly excited polaritonic modes are computed to be 
\begin{widetext}
\begin{equation}
\begin{pmatrix}\bra{2_{0}}\\
\bra{1_{0}1_{\textrm{A}}}\\
\bra{1_{\textrm{A}}^{2}}\\
\bra{2_{\textrm{A}}}
\end{pmatrix}\ket{2_{\pm}}^{(1)}=\frac{h_{\mp}^{2}}{2N\Omega_{10}}\begin{pmatrix}{{h_{+}}^{2}{h_{-}}^{2}\left[{\pm3\chi{\omega_{10}}+2\left({{h_{\mp}}^{2}-5{h_{\pm}}^{2}}\right)\gamma{\Omega_{10}}}\right]}\\
{\sqrt{2}{h_{+}}{h_{-}}\left[{\left({2{h_{\mp}}^{2}-{h_{\pm}}^{2}}\right)\chi{\omega_{10}}\pm2\left({{h_{\mp}}^{2}+2{h_{\pm}}^{2}-6{h_{+}}^{2}{h_{-}}^{2}}\right)\gamma{\Omega_{10}}}\right]}\\
{{h_{\mp}}^{2}\sqrt{\frac{{N-1}}{N}}\left[{\left({\frac{{{h_{-}}^{2}-{h_{+}}^{2}\pm3{h_{+}}^{2}{h_{-}}^{2}}}{{{h_{\pm}}^{2}}}}\right)\chi{\omega_{10}}+2\left({2{h_{\pm}}^{2}-{h_{\mp}}^{2}-6{h_{+}}^{2}{h_{-}}^{2}}\right)\gamma{\Omega_{10}}}\right]}\\
{\frac{1}{{\sqrt{N}}}\left\{ {\mp\frac{{N+{h_{\mp}}^{2}\left({{h_{\pm}}^{2}-{h_{\mp}}^{2}-3{h_{+}}^{2}{h_{-}}^{2}}\right)}}{{{h_{\pm}}^{2}}}\chi{\omega_{10}}+2\left[{N+2{h_{\mp}}^{2}\left({2{h_{\pm}}^{2}-{h_{\mp}}^{2}-6{h_{+}}^{2}{h_{-}}^{2}}\right)}\right]\gamma{\Omega_{10}}}\right\} }
\end{pmatrix}.
\end{equation}
For non-zero detuning, the corrections to the remaining totally-symmetric
eigenstates are 
\begin{align}
\begin{pmatrix}\bra{2_{0}}\\
\bra{1_{0}1_{\textrm{A}}}\\
\bra{1_{\textrm{A}}^{2}}\\
\bra{2_{\textrm{A}}}
\end{pmatrix}\ket{1_{+}1_{-}}^{(1)}= & \frac{2h_{+}h_{-}}{N\Omega_{10}}\begin{pmatrix}{-\sqrt{2}{h_{+}}^{2}{h_{-}}^{2}\gamma{\Omega_{10}}}\\
{{h_{+}}{h_{-}}\left[{\chi{\omega_{10}}-\left({{h_{+}}^{2}-{h_{-}}^{2}}\right)\gamma{\Omega_{10}}}\right]}\\
{\sqrt{\frac{{2\left({N-1}\right)}}{N}}{h_{+}}^{2}{h_{-}}^{2}\left({\frac{{2}}{{{h_{+}}^{2}-{h_{-}}^{2}}}\chi{\omega_{10}}-3\gamma{\Omega_{10}}}\right)}\\
{-\frac{1}{{\sqrt{2N}}}\left[{\frac{{N-4{h_{+}}^{2}{h_{-}}^{2}}}{{{h_{+}}^{2}-{h_{-}}^{2}}}\chi{\omega_{10}}+\left({N-6{h_{+}}^{2}{h_{-}}^{2}}\right)\gamma{\Omega_{10}}}\right]}
\end{pmatrix}
\intertext{and}\begin{pmatrix}\bra{2_{0}}\\
\bra{1_{0}1_{\textrm{A}}}\\
\bra{1_{\textrm{A}}^{2}}\\
\bra{2_{\textrm{A}}}
\end{pmatrix}\ket{2_{D\textrm{A}}}^{(1)}= & \frac{\sqrt{N-1}}{2N\Omega_{10}}\begin{pmatrix}{\frac{1}{{{h_{-}}^{2}-{h_{+}}^{2}}}\chi{\omega_{10}}}\\
{\frac{{\sqrt{2}}}{{{h_{+}}{h_{-}}}}\chi{\omega_{10}}}\\
{\sqrt{\frac{{N-1}}{N}}\left({\frac{{1-5{h_{-}}^{2}{h_{+}}^{2}}}{{{h_{+}}^{2}{h_{-}}^{2}\left({{h_{-}}^{2}-{h_{+}}^{2}}\right)}}\chi{\omega_{10}}-2\gamma{\Omega_{10}}}\right)}\\
{\frac{1}{{\sqrt{N}}}\left({\frac{{1-5{h_{-}}^{2}{h_{+}}^{2}}}{{{h_{+}}^{2}{h_{-}}^{2}\left({{h_{-}}^{2}-{h_{+}}^{2}}\right)}}\chi{\omega_{10}}-2\gamma{\Omega_{10}}}\right)}
\end{pmatrix}.
\end{align}
\end{widetext}

Under resonant conditions, these two states are degenerate in the
harmonic limit. After lifting the degeneracy, to zeroth order, these
states become 
\begin{align}
\ket{1_{+}1_{-}}^{(0)}= & \frac{\sqrt{N-1}\ket{2_{0}}-\sqrt{N}\ket{1_{\textrm{A}}^{2}}}{\sqrt{2N-1}}\\
\ket{2_{D\textrm{A}}}^{(0)}= & -\frac{\sqrt{N}\ket{2_{0}}+\sqrt{N-1}\ket{1_{\textrm{A}}^{2}}}{\sqrt{2N\qty(2N-1)}}\nonumber \\
 & +\sqrt{\frac{2N-1}{2N}}\ket{2_{\textrm{A}}}.
\end{align}
After this procedure, there are no first-order corrections for the
state $\ket{1_{+}1_{-}}$; on the other hand, 
\begin{equation}
\begin{split}\ket{2_{D\textrm{A}}}^{(1)}= & \frac{\sqrt{2N-1}}{2N\Omega_{10}}\left(-\frac{\gamma\Omega_{10}}{\sqrt{2}}\ket{2_{0}}+\chi\omega_{10}\ket{1_{0}1_{\textrm{A}}}\right.\\
 & \left.-\gamma\Omega_{10}\frac{\sqrt{N-1}\ket{1_{\textrm{A}}^{2}}+\ket{2_{\textrm{A}}}}{\sqrt{2N}}\right).
\end{split}
\end{equation}

For states carrying the standard irrep, the corrections of first-order in anharmonicity
are 
\begin{widetext}
\begin{align}
\begin{pmatrix}\bra{1_{0}1_{\textrm{B}(k)}}\\
\bra{1_{\textrm{B}(k)}^{2}}\\
\bra{2_{\textrm{B}(k)}}
\end{pmatrix}\ket{1_{\pm}1_{\textrm{B}(k)}}^{(1)}= & \frac{\sqrt{2}h_{\mp}}{N\Omega_{10}}\begin{pmatrix}{\sqrt{2}{h_{+}}{h_{-}}\left[{\chi{\omega_{10}}-\left({{h_{+}}^{2}-{h_{-}}^{2}}\right)\gamma{\Omega_{10}}}\right]}\\
{{h_{\mp}}^{2}\sqrt{\frac{{2\left({N-2}\right)}}{N}}\left[{\pm\left({\frac{{2{h_{\pm}}^{2}+{h_{\mp}}^{2}}}{{{h_{\pm}}^{2}}}}\right)\chi{\omega_{10}}-\left({3{h_{\pm}}^{2}+{h_{\mp}}^{2}}\right)\gamma{\Omega_{10}}}\right]}\\
{\frac{1}{{\sqrt{N}}}\left\{ {\mp\left[{\frac{{N-2{h_{\mp}}^{2}\left({3{h_{\pm}}^{2}+{h_{\mp}}^{2}}\right)}}{{{h_{\pm}}^{2}}}}\right]\chi{\omega_{10}}+\left[{N-2{h_{\mp}}^{2}\left({3{h_{\pm}}^{2}+{h_{\mp}}^{2}}\right)}\right]\gamma{\Omega_{10}}}\right\} }
\end{pmatrix},\intertext{and}\begin{pmatrix}\bra{1_{0}1_{\textrm{B}(k)}}\\
\bra{1_{\textrm{B}(k)}^{2}}\\
\bra{2_{\textrm{B}(k)}}
\end{pmatrix}\ket{2_{D\textrm{B}(k)}}^{(1)}= & \frac{\sqrt{2(N-2)}}{h_{+}h_{-}N\Omega_{10}}\begin{pmatrix}{\chi{\omega_{10}}}\\
{-\sqrt{\frac{{N-2}}{N}}\left({\frac{{{h_{+}}^{2}-{h_{-}}^{2}}}{{{h_{+}}{h_{-}}}}\chi{\omega_{10}}+{h_{+}}{h_{-}}\gamma{\Omega_{10}}}\right)}\\
{-\sqrt{\frac{2}{N}}\left({\frac{{{h_{+}}^{2}-{h_{-}}^{2}}}{{{h_{+}}{h_{-}}}}\chi{\omega_{10}}+{h_{+}}{h_{-}}\gamma{\Omega_{10}}}\right)}
\end{pmatrix}.
\end{align}
\end{widetext}

\begin{table*}
\caption{Eigenfrequencies in the doubly excited manifold. The Tavis-Cummings (TC) model is effectively an anharmonic system with $\chi=0$ and $\gamma=-1$; therefore, when $N\gg1$, the TC model yields the same results as the Harmonic limit.\label{tab:freqs}}

\begin{ruledtabular}
\begin{tabular}{ccc}
Eigenmode &Harmonic limit&$O(\chi,\gamma)$ anharmonic corrections\tabularnewline
\hline 
$2_{\pm}$ & $2\omega_{\pm}$&${\displaystyle -\frac{h_{\mp}^{4}}{N}\qty(\chi\omega_{10}\mp4h_{\pm}^{2}\gamma\Omega_{10})}$\tabularnewline
 &  &  \tabularnewline
$1_{+}1_{-}$ & $\omega_0+\omega_{10}$ & ${\displaystyle -\frac{2h_{+}^{2}h_{-}^{2}}{N}\qty[\qty(1-\delta_{h_{+}h_{-}})\chi\omega_{10}-2\qty(h_{+}^{2}-h_{-}^{2})\gamma\Omega_{10}]}$\tabularnewline
 &  &  \tabularnewline
$2_{D\textrm{A}}$ & $2\omega_{10}$ & ${\displaystyle -\frac{2N-2+\delta_{h_{+}h_{-}}}{2N}\chi\omega_{10}}$\tabularnewline
 &  &  \tabularnewline
$1_{\pm}1_{\textrm{B}}$ & $\omega_{10}+\omega_{\pm}$ & ${\displaystyle -\frac{2h_{\mp}^{2}}{N}\qty(\chi\omega_{10}\mp2h_{\pm}^{2}\gamma\Omega_{10})}$\tabularnewline
 &  &  \tabularnewline
$2_{D\textrm{B}}$ & $2\omega_{10}$ & ${\displaystyle -\frac{N-2}{N}\chi\omega_{10}}$\tabularnewline
 &  &  \tabularnewline
$1_{\textrm{C}}^{2}$ & $2\omega_{10}$ & 0 \tabularnewline
\end{tabular}
\end{ruledtabular}

\end{table*}

\begin{table*}
\caption{Dipolar transition intensities between eigenstates carrying the totally
symmetric irrep in the first and second excited manifolds.\label{tab:intensitiesA}}

\begin{ruledtabular}
\begin{tabular}{cccc}
$\ket{\Psi}$ & \multicolumn{3}{c}{$\mel{\Psi}{\hat{\boldsymbol{\mu}}/\mu_{10}}{\Phi}$}\tabularnewline
\hline 
 & \multicolumn{3}{c}{$\ket{\Phi}$}\tabularnewline
 &  & $\ket{1_{+}}$ & $\ket{1_{-}}$\tabularnewline
\hline 
\multicolumn{4}{c}{Tavis-Cummings}\tabularnewline
\hline 
$\ket{2_{+}}$ &  & $\sqrt{2}h_{-}\qty(h_{+}^{2}\sqrt{N}+h_{-}^{2}\sqrt{N-1})$ & $\sqrt{2}h_{+}h_{-}^{2}\qty(\sqrt{N-1}-\sqrt{N})$\tabularnewline
$\ket{2_{-}}$ &  & $\sqrt{2}h_{+}^{2}h_{-}\qty(\sqrt{N-1}-\sqrt{N})$ & $\sqrt{2}h_{+}\qty(h_{+}^{2}\sqrt{N-1}+\sqrt{2}h_{-}^{2})$\tabularnewline
$\ket{1_{+}1_{-}}$ &  & $h_{+}\qty[h_{+}^{2}\sqrt{N}+h_{-}^{2}\qty(2\sqrt{N-1}-\sqrt{N})]$ & $h_{-}\qty[h_{+}^{2}\qty(2\sqrt{N-1}-\sqrt{N})+h_{-}^{2}\sqrt{N}]$\tabularnewline
\hline 
\multicolumn{4}{c}{Harmonic approximation}\tabularnewline
\hline 
$\ket{2_{+}}$ &  & $\sqrt{2N}h_{-}$ & 0\tabularnewline
$\ket{2_{-}}$ &  & 0 & $\sqrt{2N}h_{+}$\tabularnewline
$\ket{1_{+}1_{-}}$ &  & $\sqrt{N}h_{+}$ & $\sqrt{N}h_{-}$\tabularnewline
$\ket{2_{D\textrm{A}}}$ &  & 0 & 0\tabularnewline
\hline 
\multicolumn{4}{c}{$O(\chi,\gamma)$ anharmonic corrections}\tabularnewline
\hline 
$\ket{2_{+}}$ &  & $\sqrt{\frac{2}{N}}h_{+}^{2}h_{-}^{3}\qty[-\frac{\chi\omega_{10}}{\Omega_{10}}+\qty(3h_{+}-h_{-}^{2})\gamma]$ & $-\frac{h_{+}h_{-}^{2}}{\sqrt{2N}}\qty[\qty(h_{+}^{2}+2h_{-}^{2})\frac{\chi\omega_{10}}{\Omega_{10}}-2\qty(h_{+}^{2}-h_{-}^{2}-2h_{+}^{2}h_{-}^{2})\gamma]$\tabularnewline
$\ket{2_{-}}$ &  & $-\frac{h_{+}^{2}h_{-}}{\sqrt{2N}}\qty[\qty(2h_{+}^{2}+h_{-}^{2})\frac{\chi\omega_{10}}{\Omega_{10}}-2\qty(h_{+}^{2}-h_{-}^{2}-2h_{+}^{2}h_{-}^{2})\gamma]$ & $\sqrt{\frac{2}{N}}h_{+}^{3}h_{-}^{2}\qty[\frac{\chi\omega_{10}}{\Omega_{10}}-\qty(h_{+}-3h_{-}^{2})\gamma]$\tabularnewline
$\ket{1_{+}1_{-}}$ & $\Delta\neq0$ & $\frac{2h_{+}h_{-}^{4}}{\sqrt{N}}\qty[\frac{\chi\omega_{10}}{\Omega_{10}}-\qty(3h_{+}^{2}-h_{-}^{2})\gamma]$ & $\frac{2h_{+}^{4}h_{-}}{\sqrt{N}}\qty[-\frac{\chi\omega_{10}}{\Omega_{10}}+\qty(h_{+}^{2}-3h_{-}^{2})\gamma]$\tabularnewline
 & $\Delta=0$ & $-\sqrt{\frac{N\qty(N-1)}{2N-1}}$ & $-\sqrt{\frac{N\qty(N-1)}{2N-1}}$\tabularnewline
$\ket{2_{D\textrm{A}}}$ & $\Delta\neq0$ & $\sqrt{\frac{N-1}{2N}}h_{-}\qty[\frac{2h_{+}^{2}-h_{-}^{2}}{h_{+}^{2}\qty(h_{+}^{2}-h_{-}^{2})}\frac{\chi\omega_{10}}{\Omega_{10}}-2\gamma]$ & $\sqrt{\frac{N-1}{2N}}h_{+}\qty[-\frac{h_{+}^{2}-2h_{-}^{2}}{h_{-}^{2}\qty(h_{+}^{2}-h_{-}^{2})}\frac{\chi\omega_{10}}{\Omega_{10}}-2\gamma]$\tabularnewline
 & $\Delta=0$ & $\sqrt{\frac{N}{2\qty(2N-1)}}+\frac{1}{2}\sqrt{\frac{2N-1}{2N}}\qty(\frac{\chi\omega_{10}}{\Omega_{10}}-\gamma)$ & $\sqrt{\frac{N}{2\qty(2N-1)}}-\frac{1}{2}\sqrt{\frac{2N-1}{2N}}\qty(\frac{\chi\omega_{10}}{\Omega_{10}}+\gamma)$\tabularnewline
\end{tabular}
\end{ruledtabular}

\end{table*}

\begin{table*}
\caption{Dipolar transition intensities between eigenstates carrying the standard
irrep in the first and second excited manifolds.\label{tab:intensitiesB}}

\begin{ruledtabular}
\begin{tabular}{ccc}
Model & $\mel{1_{\pm}1_{\textrm{B}}}{\hat{\boldsymbol{\mu}}/\mu_{10}}{1_{\textrm{B}}}$ & $\mel{2_{D\textrm{B}}}{\hat{\boldsymbol{\mu}}/\mu_{10}}{1_{\textrm{B}}}$\tabularnewline
\hline 
{Tavis-Cummings} & $h_{\mp}\sqrt{N-2}$ & -\tabularnewline
{Harmonic approximation} & $\sqrt{N}$ & 0\tabularnewline
{$O(\chi,\gamma)$ anharmonic corrections} & $\mp\frac{2h_{\pm}h_{+}h_{-}}{\sqrt{N}}\qty[\frac{\chi\omega_{10}}{\Omega_{10}}-\qty(h_{+}^{2}-h_{-}^{2})\gamma]$ & $-\sqrt{\frac{2\qty(N-2)}{N}}\qty(\frac{h_{+}^{2}-h_{-}^{2}}{h_{+}^{2}h_{-}^{2}}\frac{\chi\omega_{10}}{\Omega_{10}}+\gamma)$\tabularnewline
\end{tabular}
\end{ruledtabular}

\end{table*}

\begin{table*}
\caption{Photon-induced transition intensities between eigenstates carrying
the totally-symmetric irrep in the first and second excited manifolds.\label{tab:intensitiesA0}}

\begin{ruledtabular}
\begin{tabular}{cccc}
$\ket{\Psi}$ & \multicolumn{3}{c}{$\mel{\Psi}{\hat{a}_{0}}{\Phi}$}\tabularnewline
\hline 
 & \multicolumn{3}{c}{$\ket{\Phi}$}\tabularnewline
 &  & $\ket{1_{+}}$ & $\ket{1_{-}}$\tabularnewline
\hline 
\multicolumn{4}{c}{Tavis-Cummings}\tabularnewline
\hline 
$\ket{2_{+}}$ &  & $\sqrt{2}h_{+}$ & 0\tabularnewline
$\ket{2_{-}}$ &  & 0 & $-\sqrt{2}h_{-}$\tabularnewline
$\ket{1_{+}1_{-}}$ &  & $-h_{-}$ & $h_{+}$\tabularnewline
\hline 
\multicolumn{4}{c}{Harmonic approximation}\tabularnewline
\hline 
$\ket{2_{+}}$ &  & $\sqrt{2}h_{+}$ & 0\tabularnewline
$\ket{2_{-}}$ &  & 0 & $-\sqrt{2}h_{-}$\tabularnewline
$\ket{1_{+}1_{-}}$ &  & $-h_{-}$ & $h_{+}$\tabularnewline
$\ket{2_{D\textrm{A}}}$ &  & 0 & 0\tabularnewline
\hline 
\multicolumn{4}{c}{$O(\chi,\gamma)$ anharmonic corrections }\tabularnewline
\hline 
$\ket{2_{+}}$ &  & $\frac{\sqrt{2}h_{+}h_{-}^{4}}{N}\qty[\frac{\chi\omega_{10}}{\Omega_{10}}-\qty(3h_{+}-h_{-}^{2})\gamma]$ & $\frac{h_{+}^{2}h_{-}^{3}}{\sqrt{2}N}\qty(-\frac{\chi\omega_{10}}{\Omega_{10}}+4h_{+}^{2}\gamma)$\tabularnewline
$\ket{2_{-}}$ &  & $-\frac{h_{+}^{3}h_{-}^{2}}{\sqrt{2}N}\qty(\frac{\chi\omega_{10}}{\Omega_{10}}+4h_{-}^{2}\gamma)$ & $\frac{\sqrt{2}h_{+}^{4}h_{-}}{N}\qty[\frac{\chi\omega_{10}}{\Omega_{10}}-\qty(h_{+}-3h_{-}^{2})\gamma]$\tabularnewline
$\ket{1_{+}1_{-}}$ & $\Delta\neq0$ & $\frac{2h_{+}^{2}h_{-}^{3}}{{N}}\qty[\frac{\chi\omega_{10}}{\Omega_{10}}-\qty(3h_{+}^{2}-h_{-}^{2})\gamma]$ & $\frac{2h_{+}^{3}h_{-}^{2}}{{N}}\qty[\frac{\chi\omega_{10}}{\Omega_{10}}-\qty(h_{+}^{2}-3h_{-}^{2})\gamma]$\tabularnewline
 & $\Delta=0$ & $-\sqrt{\frac{N-1}{2N-1}}$ & $-\sqrt{\frac{N-1}{2N-1}}$\tabularnewline
$\ket{2_{D\textrm{A}}}$ & $\Delta\neq0$ & $-\frac{1}{N}\sqrt{\frac{N-1}{2}}\frac{h_{-}^{2}}{h_{+}\qty(h_{+}^{2}-h_{-}^{2})}\frac{\chi\omega_{10}}{\Omega_{10}}$ & $\frac{1}{N}\sqrt{\frac{N-1}{2}}\frac{h_{+}^{2}}{h_{-}\qty(h_{+}^{2}-h_{-}^{2})}\frac{\chi\omega_{10}}{\Omega_{10}}$\tabularnewline
 & $\Delta=0$ & $-\frac{1}{\sqrt{2\qty(2N-1)}}+\frac{1}{2N}\sqrt{\frac{2N-1}{2}}\qty(\frac{\chi\omega_{10}}{\Omega_{10}}-\gamma)$ & $\frac{1}{\sqrt{2\qty(2N-1)}}+\frac{1}{2N}\sqrt{\frac{2N-1}{2}}\qty(\frac{\chi\omega_{10}}{\Omega_{10}}+\gamma)$\tabularnewline
\end{tabular}
\end{ruledtabular}

\end{table*}

\begin{table*}
\caption{Photon-induced transition intensities between eigenstates carrying
the standard irrep in the first and second excited manifolds.\label{tab:intensitiesB0}}

\begin{ruledtabular}
\begin{tabular}{ccc}
Model & $\mel{1_{\pm}1_{\textrm{B}}}{\hat{a}_{0}}{1_{\textrm{B}}}$ & $\mel{2_{D\textrm{B}}}{\hat{a}_{0}}{1_{\textrm{B}}}$\tabularnewline
\hline 
{Tavis-Cummings} & $\pm h_{\pm}$ & -\tabularnewline
{Harmonic approximation} & $\pm h_{\pm}$ & 0\tabularnewline
{$O(\chi,\gamma)$ anharmonic corrections} & $\mp\frac{2h_{\pm}h_{+}h_{-}}{\sqrt{N}}\qty[\frac{\chi\omega_{10}}{\Omega_{10}}-\qty(h_{+}^{2}-h_{-}^{2})\gamma]$ & $-\sqrt{\frac{2\qty(N-2)}{N}}\qty(\frac{h_{+}^{2}-h_{-}^{2}}{h_{+}^{2}h_{-}^{2}}\frac{\chi\omega_{10}}{\Omega_{10}}+\gamma)$\tabularnewline
\end{tabular}
\end{ruledtabular}

\end{table*}

The main lesson from these perturbative correction calculations is
that the anharmonicity-induced energy shifts should be undetectable
as they scale as $1/N\to0$ when $N\gg1$, as is the case of collective
SC. Hence, calculations using one or small number of emitters cannot
accurately model the nonlinear response of vibrational polariton systems
under those conditions.

\section{Conclusions \label{sec:conc}}

In this work, we have found the exact eigenspectrum of the doubly
excited manifold of an ensemble of $N$ anharmonic oscillators under
collective VSC. We provided a group-theoretical formalism to solve
this many-body problem, reducing it to small matrix diagonalizations
(the largest of which involves a $4\times4$ matrix). This procedure
is a significant simplification from the brute-force numerical approach,
which involves diagonalization of astronomically large ${N+2 \choose 2}$-dimensional matrices where $N=10^6-10^{10}$. We provide compact expressions and tables that should
serve as a concise reference for future work involving the nonlinear spectroscopy
of molecular polariton systems.

Through numerically exact examples and analytical studies using perturbation
theory, we have contrasted this model to the standard TC and HO
models, and have shown that the additionally available anharmonic transitions per molecule  can
give rise to new phenomena, such as enhancement of two-photon absorption
cross-sections owing to new resonances between bipolariton states
and anharmonic two-quanta vibrational states. We have demonstrated
that there are at most two such bipolariton states that fullfill those
resonant conditions. These conclusions, while studied specifically
for vibrational SC conditions, should have analogues in other frequency ranges,
such as in the UV-visible regime, where anharmonic shifts in electronic
transitions are ubiquitous.

We have provided with a summary of analytical expresions for eigenfrequencies and matrix elements of typical transition operators. These might serve as a reference guide for future theoretical and experimental explorations of the doubly excited manifold. We highlight the finding that the deviation from the harmonic behavior is negliglible for large $N$. This statement seems to contradict reports of strong optical nonlinearities in systems under collective SC.\cite{Dunkelberger2016,Xiang2018} The resolution to this apparent conundrum lies in recognizing the essential role of anharmonic dissipative processes in the creation of reservoirs populated with dark states.\cite{Ribeiro2018} This effect is not considered in the present formalism as it ignores dissipation, and focuses on the regime where multiphoton absorption processes occur faster than relaxation into dark states.  

Finally, our work also shows that calculations involving few emitters
and higher excitation manifolds (i.e. nonlinear response calculations) cannot provide
a satisfactory approximation to the nonlinear optics of the collective SC regime even if
the single-molecule dipole is artificially increased to keep the overall
light-matter coupling constant. We hope that these findings inform
future experimental endeavors and the theoretical considerations regarding
polaritonic chemical behavior and non-linear response.

\appendix

\section{Exact expressions for TC solutions.\label{sec:TCexact}}

Diagonalization of the TC Hamiltonian carrying the totally-symmetric
irrep, \cref{eq:tca}, entails solving a cubic characteristic polynomial.
Although it is possible to obtain analytical expressions, this task
is rather complicated and a numerical approach can be easily implemented
if exact solutions are needed. Nonetheless, we list the eigenfrequencies
for the corresponding bipolaritons as a reference: 
\begin{equation}
\ev{\hat{H}_{2}}{\Psi}_{\textrm{TC}}=\omega_0+\omega_{10}+f_{\Psi}(\rho)\Omega_{10},
\end{equation}
where \begin{subequations}\label{eq:exacttca} 
\begin{align}
f_{2+}(\rho)= & 2\rho',\\
f_{2-}(\rho)= & -\rho'-\sqrt{3}\rho'',\intertext{and}f_{1+1-}(\rho)= & -\rho'+\sqrt{3}\rho''.
\end{align}
\end{subequations} In \cref{eq:exacttca}, $\rho'$ and $\rho''$
refer to the real and imaginary parts, respectively, of the quantity
$\rho$, which fulfills \begin{subequations} 
\begin{equation}
\rho^{3}=q^{3}+\textrm{i}\sqrt{p^{6}-q^{6}},
\end{equation}
where the coefficients $p$ and $q$ are such that 
\begin{align}
p^{2}=\frac{N-2h_{+}^{2}h_{-}^{2}}{3N}=\frac{1}{3}\qty(1-\frac{2g_{10}^{2}}{\Omega_{10}^{2}}),\intertext{and}q^{3}=\frac{h_{+}^{2}h_{-}^{2}}{N}\qty(h_{+}^{2}-h_{-}^{2})=\frac{g_{10}^{2}}{\Omega_{10}^{3}}\Delta.
\end{align}
\end{subequations}

Finally, the eigenvectors are given by 
\begin{equation}
\ket{\Psi}=\mathcal{N}_{\Psi}\begin{pmatrix}2\sqrt{N-1}h_{+}^{2}h_{-}^{2}\\
\sqrt{2\qty(N-1)}h_{+}h_{-}\varphi_{\Psi}(\rho)\\
\sqrt{N}\qty(f_{\Psi}(\rho)\varphi_{\Psi}(\rho)-2h_{+}^{2}h_{-}^{2})
\end{pmatrix},
\end{equation}
where $\varphi_{\Psi}(\rho)=f_{\Psi}(\rho)-h_{+}^{2}+h_{-}^{2}$,
and $\mathcal{N}_{\Psi}$ is a normalization coefficient.

\section{Fourier basis with two excitations.\label{sec:four2x}}

In this section, we discuss the dark states that emerge in the doubly
excited manifold beyond of what was covered on the main body of the
manuscript. In the harmonic limit, a transparent way to generate
dark states with meaningful labels is through the application of two
creation operators related to the eigenmodes. Explicitly, these states
are \begin{subequations}\label{eq:darknonsym} 
\begin{align}
\ket{2_{\gamma D(k)}} & =\frac{\hat{a}_{\textrm{B}(k)}^{\dagger}}{\sqrt{2}}\ket{1_{\textrm{B}(k)}},\intertext{and}\ket{1_{\gamma D(k\ell)}^{2}} & =\hat{a}_{\textrm{B}(k)}^{\dagger}\ket{1_{\textrm{B}(\ell)}}=\hat{a}_{\textrm{B}(\ell)}^{\dagger}\ket{1_{\textrm{B}(k)}}.
\end{align}
\end{subequations} The labels $k$ and $\ell$ clearly indicate either
the emitters in which the wavefunction is mostly localized, if the
Schur-Weyl basis is used, or the wave numbers assigned to the eigenfunction,
in the case of the Fourier basis. However, these functions no longer
carry a defined irrep, but instead the $N\qty(N-1)/2$-dimensional
reducible representation $\gamma=\textrm{A}+\textrm{B}+\textrm{C}$ (not to be confused with the electric anharmonicity parameter). 

The generation of states carrying the standard irrep has been presented
in section \ref{sec:2man}, as well as the explicit form of the C-symmetric
states in the Schur-Weyl basis, where 
\begin{equation}
\alpha_{mn}^{(k\ell)}=\alpha_{m}^{(k)}\alpha_{n}^{(\ell)}+\alpha_{n}^{(k)}\alpha_{m}^{(\ell)}-2\qty(\alpha_{m}^{(k)}\delta_{n\ell}+\alpha_{n}^{(k)}\delta_{m\ell}).
\end{equation}

In the case of the Fourier basis, we illustrate the application of
\cref{eq:4drk} for a scenario with four emitters. After removing
the functions with symmetries A and B, for $N=4$, we have $N\qty(N-3)/2=2$,
and it can be checked, with the Schur-Weyl basis, that the allowed
values for $k$ and $\ell$ are $\{2,3\}$ and $4$, respectively.
For $N=4$, we can define the wave vectors proportional to $0\leq w\leq3$.
This fact gives rise to the combinations of center of mass, $K$,
and relative wave number, $q$, shown in \cref{tab:kandq}. 
\begin{table}
\caption{Coefficients of center of mass and relative wave number for wavefunctions
with $N=4$.\label{tab:kandq}}

\begin{ruledtabular}
\begin{tabular}{lccccr}
 & $w$ & $w'$ & $K$ & $q$ & \tabularnewline
\hline 
 & 0 & 0 & 0 & 0 & \tabularnewline
 & 0 & 1 & 1 & 1 & \tabularnewline
 & 0 & 2 & 2 & 2 & \tabularnewline
 & 0 & 3 & 3 & 3 & \tabularnewline
 & 1 & 1 & 2 & 0 & \tabularnewline
 & 1 & 2 & 3 & 1 & \tabularnewline
 & 1 & 3 & 4 & 2 & \tabularnewline
 & 2 & 2 & 4 & 0 & \tabularnewline
 & 2 & 3 & 5 & 1 & \tabularnewline
 & 3 & 3 & 6 & 0 & \tabularnewline
\end{tabular}
\end{ruledtabular}

\end{table}

Plugging these into \cref{eq:4drk}, we get 
\begin{widetext}
\begin{multline}
c_{mn}^{(k,4)}=\psi(0,0)+\psi(2,0)\textrm{e}^{\frac{\pi i}{2}\qty(m+n)}+\psi(4,0)\textrm{e}^{\pi i\qty(m+n)}+\psi(6,0)\textrm{e}^{\frac{3\pi i}{2}\qty(m+n)}\\
+\qty[\psi(1,1)\textrm{e}^{\frac{\pi i}{4}\qty(m+n)}+\psi(3,1)\textrm{e}^{\frac{3\pi i}{4}\qty(m+n)}+\psi(5,1)\textrm{e}^{\frac{5\pi i}{4}\qty(m+n)}]\cos\qty(\pi\frac{m-n}{4})\\
+\qty[\psi(2,2)\textrm{e}^{\frac{\pi i}{2}\qty(m+n)}+\psi(4,2)\textrm{e}^{\pi i\qty(m+n)}]\cos\qty(\pi\frac{m-n}{2})+\psi(3,3)\textrm{e}^{\frac{3\pi i}{4}\qty(m+n)}\cos\qty(3\pi\frac{m-n}{4}).
\end{multline}
\end{widetext}

With help of Gram-Schmidt orthogonalization, it is possible to conclude
that all coefficients $\psi(K_{k,4},q_{k,4})$ vanish, except for $\psi(6,0)$,
$\psi(4,0)$, $\psi(4,2)$, and $\psi(2,2)$. With this information, we can write
write \begin{subequations}\label{eq:cfin}
\begin{align}
c_{mn}^{(k,4)}= & \frac{1}{2\sqrt{3}}\textrm{e}^{\pi i\qty(m+n)}\qty[\cos\qty(\pi\frac{m-n}{2})-1],\intertext{and}c_{mn}^{(k',4)}= & -\frac{1}{2}\qty[\textrm{e}^{\frac{3\pi i}{2}\qty(m+n)}+\textrm{e}^{\frac{\pi i}{2}\qty(m+n)}\cos\qty(\pi\frac{m-n}{2})].
\end{align}
\end{subequations} 
Identifying the appropriate values for $k$ and $k'$ is rather cumbersome and escapes the scope of this work.

\begin{acknowledgements}
JACGA thanks Stephan van den Wildenberg and Matthew Du for their useful comments and insights. The authors acknowledge funding support from the Air Force Office of Scientific Research award FA9550-18-1-0289.
\end{acknowledgements}

\bibliographystyle{aip}
\bibliography{2exmanbib}

\end{document}